\documentclass[a4paper,fleqn]{cas-sc}

\usepackage[authoryear]{natbib}
\usepackage{geometry}
\usepackage{gensymb}
\usepackage{graphicx}
\usepackage{subcaption}
\usepackage{amsmath}
\usepackage{lineno}

\usepackage{color}
\usepackage{soul}

\def\tsc#1{\csdef{#1}{\textsc{\lowercase{#1}}\xspace}}
\tsc{WGM}
\tsc{QE}
\begin{document}
\let\WriteBookmarks\relax
\def\floatpagepagefraction{1}
\def\textpagefraction{.001}

\shorttitle{Subsurface dynamics in oblique granular impacts}    

\shortauthors{Miklav\v ci\v c, S\'anchez, Wright, Quillen, Askari}  

\title[mode = title]{Sub-surface granular dynamics in the context of oblique, low-velocity impacts into angular granular media}  

%\tnotemark[<tnote number>] 

%\tnotetext[<tnote number>]{<tnote text>} 


\author[1]{Peter M. Miklav\v ci\v c}[orcid=0000-0002-9724-8744]

\cormark[1]


\ead{pmiklavc@ur.rochester.edu}

%\credit{<Credit authorship details>}

\affiliation[1]{organization={Department of Mechanical Engineering, University of Rochester},
            street={235 Hopeman Building, P.O. Box 270132,}, 
            city={Rochester},
            state={NY},
            postcode={14627}, 
            country={USA}}
            
\affiliation[2]{organization={Colorado Center for Astrodynamics Research, University of Colorado Boulder},
            street={3775 Discovery Dr.,},
            city={Boulder},
            state={CO},
            postcode={80303},
            country={USA}}
            
\affiliation[3]{organization={Department of Physics and Astronomy, University of Rochester},
            street={206 Bausch and Lomb Hall, P.O. Box 270171,}, 
            city={Rochester},
            state={NY},
            postcode={14627}, 
            country={USA}}

\author[2]{Paul S\'anchez}
\ead{diego.sanchez-lana@colorado.edu}

\author[3]{Esteban Wright}
\ead{ewrig15@ur.rochester.edu}

\author[3]{Alice C. Quillen}
\ead{alice.quillen@rochester.edu}

\author[1]{Hesam Askari}
\ead{askari@rochester.edu}

% % Credit authorship

% % Address/affiliation
% \affiliation[<aff no>]{organization={},
% %          citysep={}, % Uncomment if no comma needed between city and postcode

% % Corresponding author text
% \cortext[1]{Corresponding author}

% % Footnote text


\begin{abstract}
Oblique, low-velocity impacts onto non-terrestrial terrain are regular occurrences during space exploration missions. These are not only a necessary component of landing and sampling maneuvers, but can also be used as impact experiments to reveal characteristics of the interacting surfaces. We conduct two-dimensional discrete simulations to model such impacts into a bed of triangular grains. Finite element method provides the basis for simulation, enabling the angular grain geometry. Our findings re-create the three classes of impact behavior previously noted from experiments: full-stop, rollout, and ricochet \citep*{Wright2020}. An application of Set Voronoi tessellation assesses packing fraction at a high resolution, revealing how grains shift relative to each other during an impact event. We also assess how packing fraction at the point of impact influences different impact behavior types. Calculation of Von Mises strain distributions then reveal how grains shift relative to the overall system, leading to the notion of the `skin zone'. Intuition would suggest that the region of perturbed grains would grow deeper with higher velocity impacts, results instead show that increasing velocity may evoke a change in the grains' dissipative response that dispatches energy predominantly laterally from the impact site instead of deeper into the bed. Finally, we consider  how sub-surface response could link with impactor dynamics to deepen our understanding of oblique, low-velocity impact events, one day helping to improve mission outcomes.
\end{abstract}


\begin{highlights}
\item Finite Element Methods support modeling of impacts into discrete triangular grains
\item Simulations reproduce experimental ricochet, roll-out, and full-stop behaviors of projectile
\item Visualization of local grain packing reveals breadth of granular disturbance from impact
\item Strain distributions reveal zones of perturbed media around the crater
\item Higher velocity impacts spur increasingly lateral granular dissipative mechanisms
\end{highlights}

% Keywords
% Each keyword is separated by \sep

\maketitle

%\linenumbers

% Main text
\section{Introduction}
\label{sec:introduction}

In recent decades, navigation of non-terrestrial terrain has presented a new front to the field of granular mechanics. Such environments differ from Earth because of different levels of gravity, low atmospheric pressure, and materials that differ from those commonly found on Earth. Asteroids are one such environment of interest. The population of Near-Earth-Asteroids within reach of spacecraft are theorized to be composed entirely of rubble \citep{Burns1975, Pravec2000, Walsh2018}. These asteroids form over time by accretion of fragments of rock held together by self-gravity, a process that has spurred simulation research of its own \citep{Leinhardt2000, Bagatin2001,  Michel2002, Richardson2005, Sanchez2011, Sanchez2014, Sanchez2015, Ferrari2019, Ferrari2020,  Michel2020, Sanchez2021}. Missions to these asteroids have returned images and even samples of the surface asteroid material, offering insight into the harsh environments that spacecraft must inevitably endure \citep{Yano2006, Tsuchiyama2011, Nakamura2011, Watanabe2017}.

Low velocity impacts on non-terrestrial surfaces have been subject of limited interest in the literature. Nonetheless, future exploration missions - to moons, planets, or asteroids - will likely encounter landings upon a rubble or regolith surface, making study of such impacts important for mission preparedness. The MASCOT lander touching down on asteroid Ryugu underlines this importance, as it has become the subject of simulations studying the impact at the landing site \citep{Thuillet2018, Thuillet2021}. Other literature studies the populations of craters observed on a planet's or asteroid's surface \citep{Gou2018} focusing on such metrics as the depth-to-diameter ratio \citep{Noguchi2021}. Morphometry of craters have been connected through various scaling research \citep{Goldman2008,Dowling2013, Celik2022}. The space shuttle and parabolic test flights have also supported experimental campaigns for studying low-velocity normal impacts into regolith simulant \citep{Brisset2018}. Oblique impacts have been less well studied experimentally, though recently Wright et al. studied impacts of a marble into sand and gravel as a function of impact angle and velocity, classifying resultant behavior into categories dependent upon whether the projectile ricocheted off the media or not \citep{Wright2020,Wright2021}. This body of work provides valuable information about the surface and overall projectile behavior in low velocity impact conditions, but the dynamics of sub-surface grains and how these correspond to the impact dynamics are, as yet, rarely studied.

Discrete element simulations offer great potential for studying sub-surface dynamics, though such methods are constrained by how accurately and efficiently each grain and its local interactions are modeled. A standard approach is to assume a spherical grain geometry which enables streamlined algorithms for contact detection and response \citep{Cundall1979}. Nonetheless, real grain geometries are never perfectly round. This could not be more true for regolith, where grains are rougher than Earthen quartz sands due to an absence of erosive forces such as water and atmosphere. Regolith also behaves differently than quartz sands. Under cyclic loading conditions, regolith has been shown to continually deform with subsequent loadings, contrasting to quartz sands that exhausts all plastic deformation within the first cycle \citep{Sandeep2019}. Given that quartz sand is often modeled using simple discrete spheres, it stands to reason that a model well-suited for simulating regolith or any irregular, coarse grain could be improved by adopting a different grain geometry. A growing segment of literature has exemplified this. The study of railroad ballast - the coarse gravel piled beneath railroad ties - adopts polyhedral grains in discrete simulation \citep{Lobo-Guerrero2006, McDowell2016, Suhr2017, Li2018}. Recent models of rubble-pile asteroids have adapted to simulate polyhedral grains instead of standard spherical grains \citep{Ferrari2019, Ferrari2020, Sanchez2021}. Also, poly-ellipsoidal grains are shown to better replicate the experimental results of a rover wheel digging in regolith simulant compared to spherical grains \citep{Knuth2012}.

Driven by the trend towards modeling angular grains, the sparsity of relevant impact literature, and the potential wealth of knowledge in observing sub-surface granular dynamics, we use Finite Element Method (FEM) - via commercial software Abaqus - to discretely simulate a two-dimensional angular grain system impacted by a disk-like projectile. We call this combination of finite element and discrete element methodologies the Discrete Finite Element Method (DFEM). Specifically, we consider media composed of mono-disperse triangular grains, a simplistic angular geometry without complexities that may introduce numerical instabilities or make models computationally inefficient. The space-filling nature of the triangular grains also supports our later study of grain packing fraction, since a much wider range of packing density is observed with triangles instead of circles. The two-dimensional basis streamlines data extraction and simplifies calculation procedures for obtaining sub-surface packing and strain metrics. These simulations offer access to bulk data for each element of the virtual system such as coordinate data or nodal fields. The packing of grains within a granular system shows how they re-arrange in response to a dynamic event \citep{Arevalo2010, Clark2012, Jia2012, Kondic2012, Xu2014, Zhao2017}. Such packings are observable experimentally with X-ray tomography \citep{Fu2006, MorenoAtanasio2010, Weis2017, Reimann2017}, though simulations allow the same perspective but with reduced hassle. We will show this using an adapted form of Voronoi tessellation \citep{Voronoi1908}. Sub-surface strain distribution caused by the surface impacts are also studied using methods applied to other granular systems in the literature \citep{Chen2011, Guo2016, Luo2017}. In summation, our methods enable a deeper level of system analysis that is unattainable with experimental methods.

The application of DFEM - FEM for discrete simulations - is present in the literature, albeit sparse. In the simulation of a two-dimensional shear cell, \cite{Kabir2008} used FEM to model discrete triangular, circular, and square grains with multiple elements per grain, visualizing force chains developed within the cell. Some angular DEM simulations have also used FEM as an intermediary between simulation steps to facilitate breakage of grains during modeling of rockfill \citep{BagherzadehKh2011, Raisianzadeh2018, Raisianzadeh2019, Seyyedan2021, Nitka2020}. The field of multi-particle finite element method (MPFEM) simulates powders discretely with individual meshes to observe powder compaction under load \citep{Guner2015, Zhang2015}.

In this paper, we present a computationally efficient method of modeling angular grains in a DEM scheme based on FEM formulation called Discrete Finite Element Method (DFEM) that allows the study of surface impacts and reveals the role of sub-surface properties on the outcome of impact events. Particularly, we present how local packing fraction can result in variations in impact characteristics and the implications of impact events on crater shape and depth as well as the affected area around the impact site. Section \ref{sec:methods} will review the simulation design, procedures for conducting impact tests, and the ranges of impact conditions evaluated. Section \ref{sec:results} will review modeling results. Finally, Section \ref{sec:conclusion} will summarize findings and re-iterate their broader implications. Appendix I showcases additional simulation work that verifies the performance of the Abaqus platform against LAMMPS DEM software \citep{LAMMPS}. Appendix II demonstrates the potential of nodal outputs from discrete models for viewing stress distributions through granular media.

\section{Simulation methods and setup}
\label{sec:methods}

Monodisperse two-dimensional discrete triangles are the base grain geometry modeled. This geometry represents a step towards the space-filling and rough character of naturally-occurring regolith grains while prioritizing model efficiency with a simple geometry. The finite-elements platform means that only one element is required per grain when formulated in Abaqus, minimizing the risk of improper deformation modes. The simplicity will also assist post-processing algorithms later on.

The simulation is comprised of a circular impactor and the triangular grains. The impactor has a diameter of 16.15mm - the same as the marble used by \cite{Wright2020} - and the edge length of the right-triangle grain is 3mm, giving an impactor-to-grain size ratio of 5.38. Both entities are modeled as elastic bodies with the material properties shown in Table \ref{tab:props}. In this work, the system is under Earth-like gravity to improve simulation efficiency (grains settle faster, reducing the overall number of increments to solve) and enable comparison to experimental literature for qualitative verification of model performance. However, this simulation setup accommodates adjustments to gravitational acceleration to study impacts in low gravity which will be explored in future.

\begin{figure}
    \centering
    \begin{subfigure}[b]{0.4\linewidth}
        \includegraphics[width=1.85in]{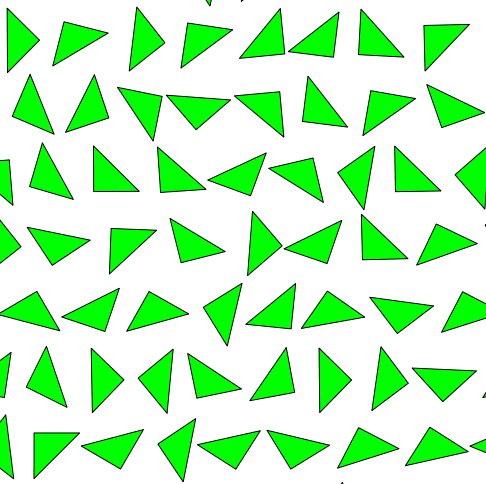}
    \end{subfigure}
    \begin{subfigure}[b]{0.4\linewidth}
        \includegraphics[width=2.0in]{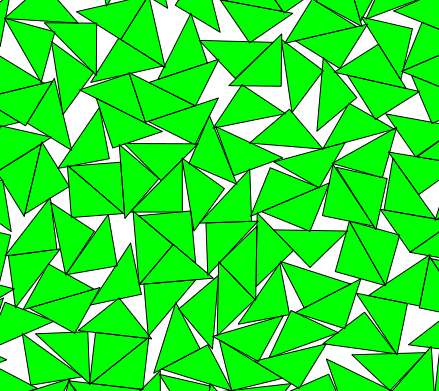}
    \end{subfigure}
    \caption{The granular bed is formed from an initial sparse patterning of randomly oriented and shifted grains (\textit{left}) that are allowed to fall and settle under gravity (\textit{right}).}
    \label{fig:grainPatterns}
\end{figure}

To set up the system for initial grain settling, a lattice grid is positioned above the rigid bin that acts as a container for granular material. A triangular grain is created and randomly positioned and oriented within each lattice cell. An external MATLAB code generates the vertex coordinates for all grains and writes those coordinates to a text file to be referenced by the Abaqus input file which will control the bed-settling stage of the simulation and results in a randomized initial configuration. The initial patterning of grains and their randomized orientation is shown in the left pane of Figure \ref{fig:grainPatterns}. Each grain was meshed as one single three-node linear plane strain element (type CPE3 in Abaqus). The impactor was discretized into thirty-two, four-node bi-linear plane strain quadrilateral elements with reduced integration (type CPE4R in Abaqus). 

\begin{table}[h!]
\centering
    \caption{Material properties used in DEM simulation.}
    \begin{tabular}{|c|c|c|c|}\hline
     & grain density & Young's & Poisson \\
     & ($kg/m^3$) & modulus ($GPA$) & ratio \\\hline
     disk & 2500 & 4.5 & 0.2 \\\hline
     grains & 2600 & 7.0 & 0.2 \\\hline
    \end{tabular}
    \label{tab:props}
\end{table}

An automatically chosen, flexible time increment was used to execute the simulations. This selection is a feature of Abaqus where a stable time increment is computed as a factor of the shortest time for a dilatational wave to cross an element in the model. To form the granular bed, the grains are allowed to fall freely and settle under Earth's gravity into the rigid bin. The tangential interaction between all of the grains as well as the impactor are defined using a friction coefficient of $\mu = 0.84$, appropriate for coarse gravel \citep{Wright2021}. Normal interactions are dictated by the material properties of the involved bodies. \citep{h98} A ramped acceleration-based shake is applied to the settled granular bed to further settle the system. The details of the vertical shake procedure are included in Table \ref{tab:gravs}. Once gravity is returned to 9.81 $m/s^2$, the simulation is allowed to run from 1.1s to 1.3s such that the total energy - calculated by Abaqus as the sum of strain energy, kinetic energy, viscous dissipation, frictional dissipation, and internal heat energy - stabilizes to a constant value.

\begin{table}
\centering
    \caption{Details of the gravity-based shake applied during the settling simulation to further settle the granular bed. Negative accelerations are directed downwards.}
    \begin{tabular}{|c|c|c|c|c|c|c|}\hline
     sim time ($s$) & 0 & 0.3 & 0.55 & 0.8 & 1.1 & 1.4 \\\hline
     gravity ($m/s^2$) & -9.81 & -9.81 & +5.0 & -12.0 & -9.81 & -9.81 \\\hline
    \end{tabular}
    \label{tab:gravs}
\end{table}

The dimensions of the granular bed are approximately 0.32 m long and 0.16 m deep. The depth is intentionally about 10 times the diameter of the impactor. The length is such that, for a 30\degree \:impact at 10 m/s, the peak reaction force felt by the rigid container is approximately 5\% of the peak reaction force felt by the projectile. This indicates the vast majority of energy is dissipated within the granular material and that effects of the rigid container on projectile response are minimal.

\begin{figure}
    \centering
    \begin{subfigure}[b]{0.49\linewidth}
        \includegraphics[width=3.1in, trim={0 0cm 0cm 0cm}, clip]{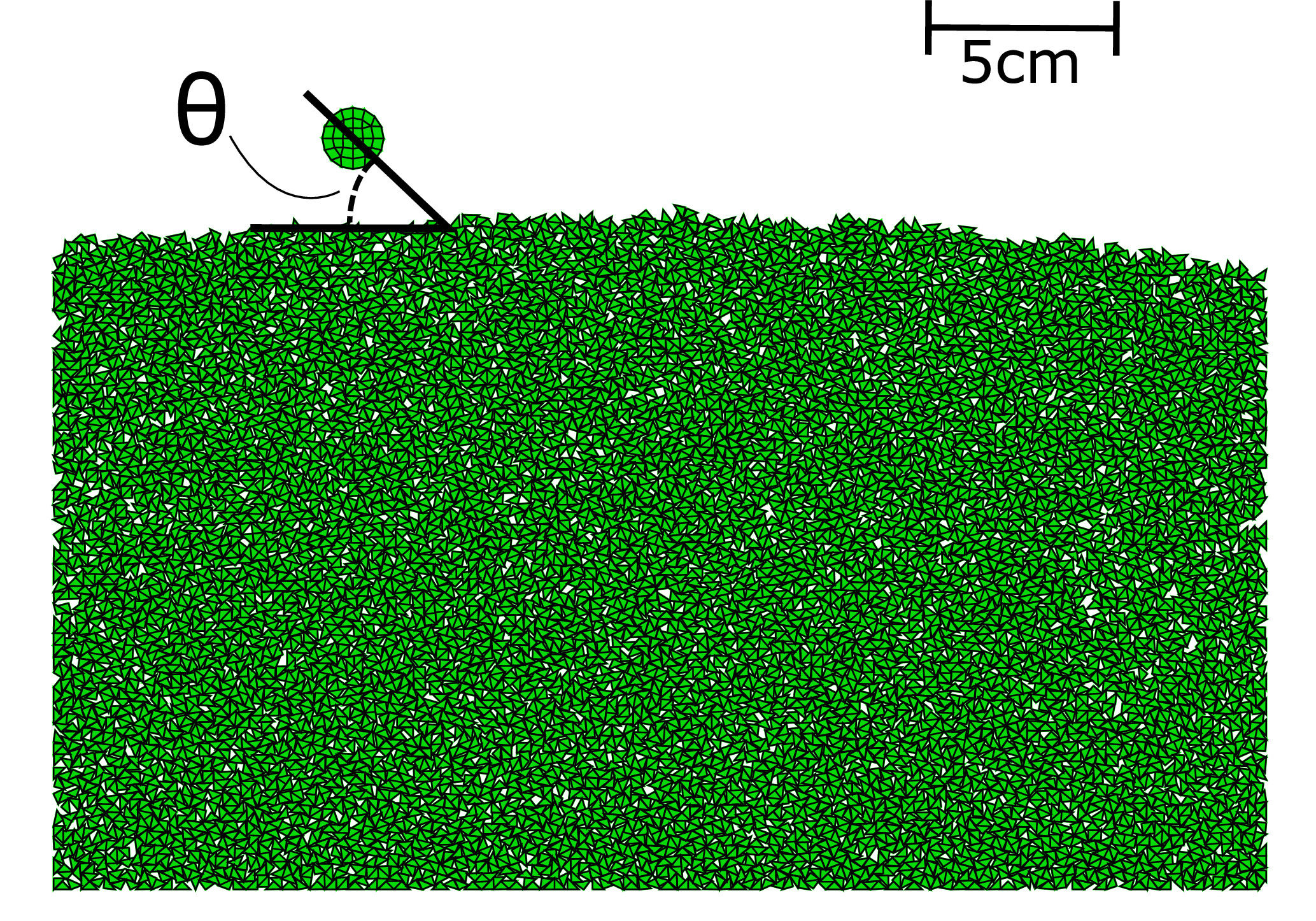}
        \subcaption{Initial configuration of disc relative to grains}
    \end{subfigure}
    \begin{subfigure}[b]{0.49\linewidth}
        \includegraphics[width=3.1in, trim={0 0cm 0cm 0cm}, clip]{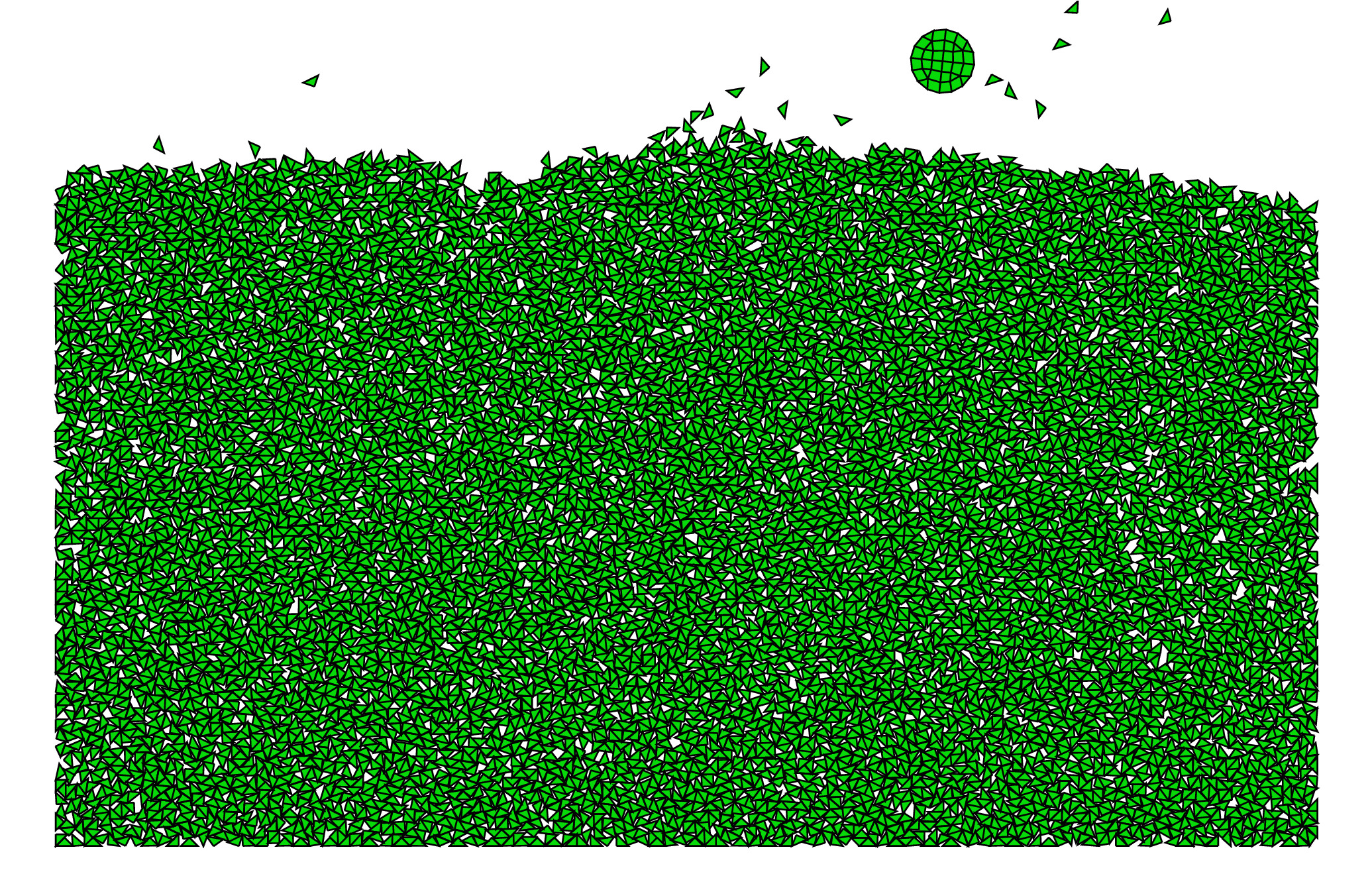}
        \subcaption{Ricochet impact event}
    \end{subfigure}
    \begin{subfigure}[b]{0.49\linewidth}
        \includegraphics[width=3.1in, trim={0 0cm 0cm 0cm}, clip]{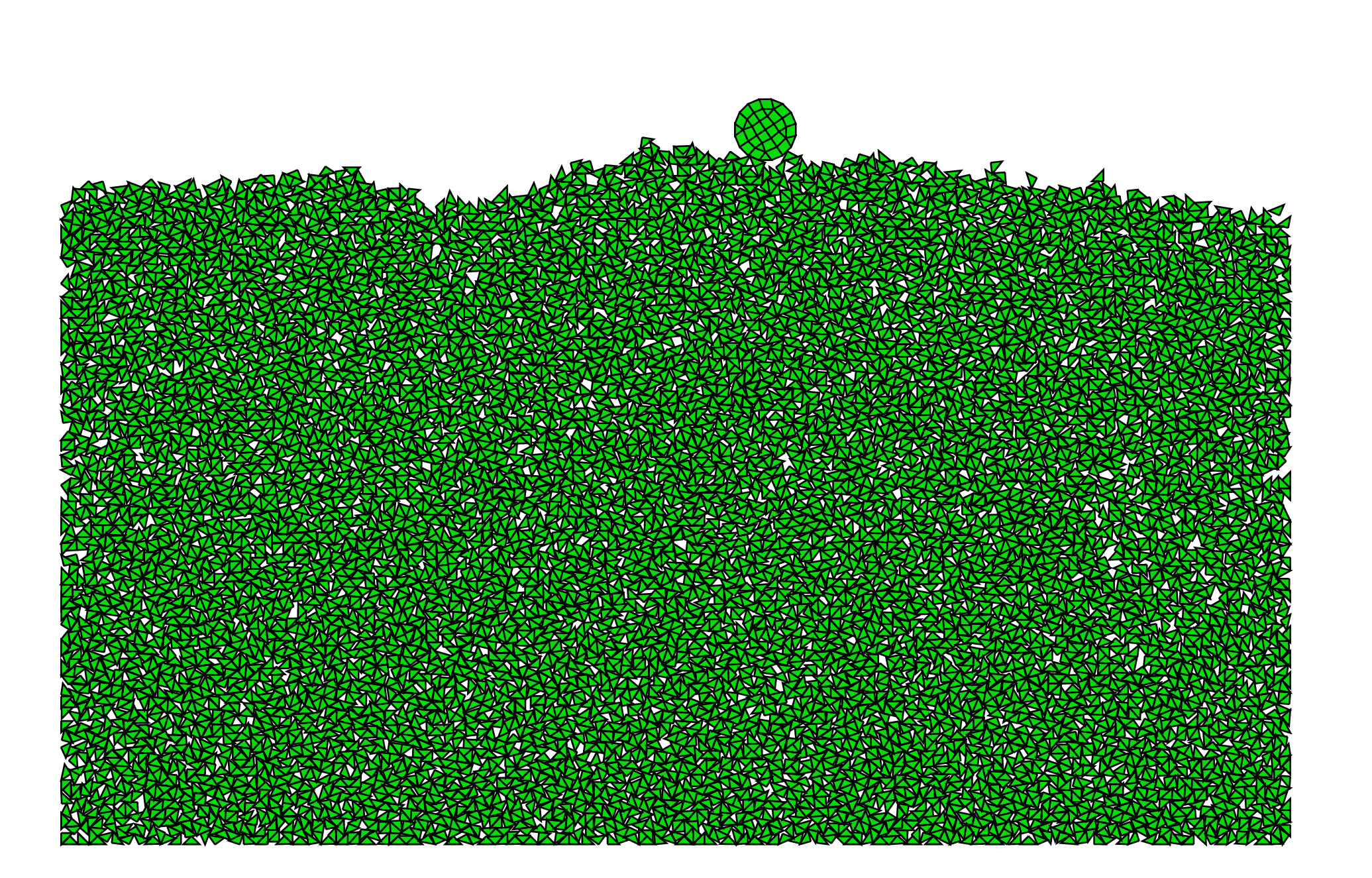}
        \subcaption{Roll-out impact event}
    \end{subfigure}
    \begin{subfigure}[b]{0.49\linewidth}
        \includegraphics[width=3.1in, trim={0 0cm 0cm 0cm}, clip]{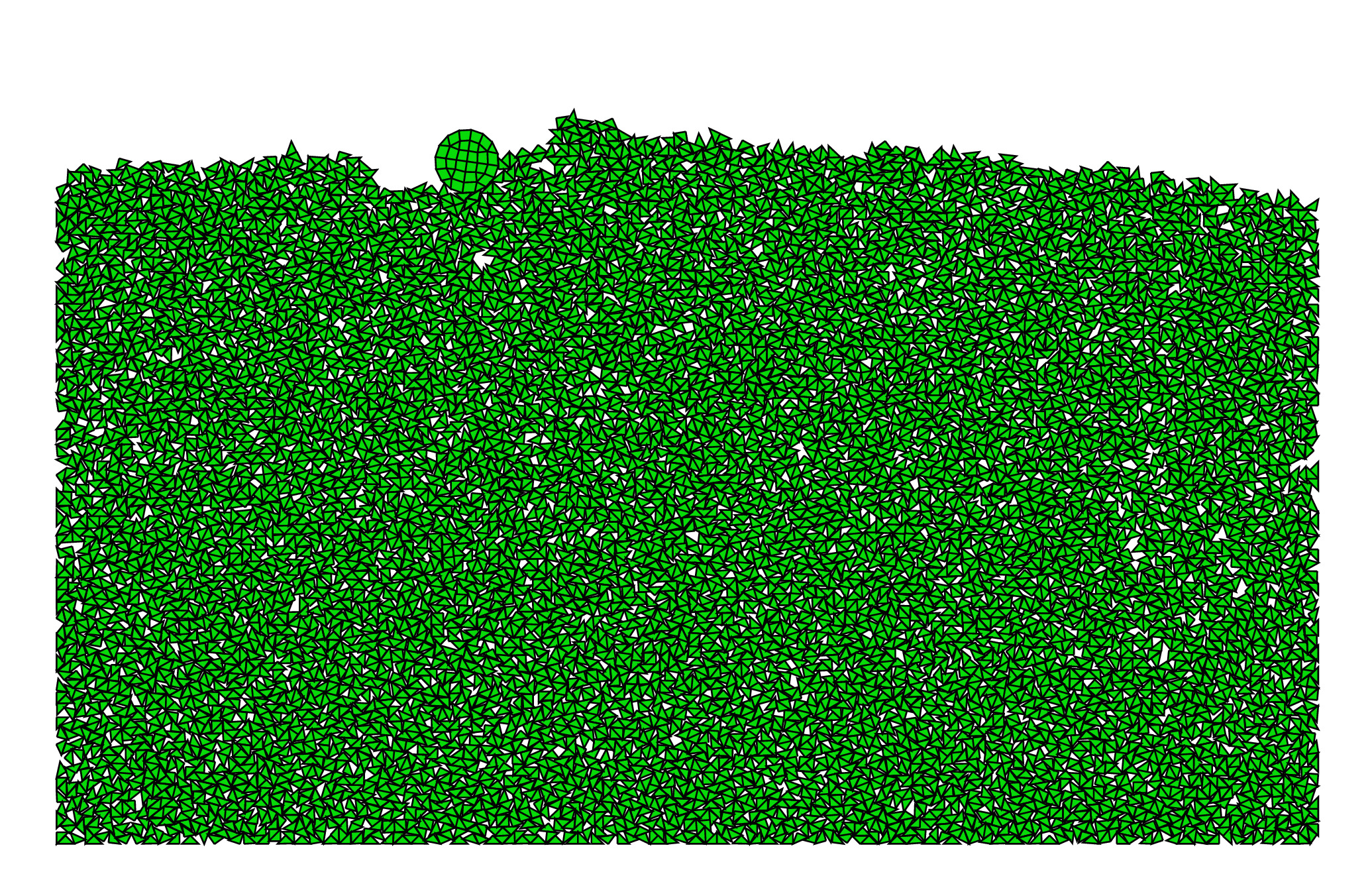}
        \subcaption{Full-stop impact event}
    \end{subfigure}
    \caption{The initial configuration of impactor and granular bed system in the DFEM model with impact angle convention (a) and final frame grabs of the three impact types (b-d). Panels (b) through (d) do not display the full depth of the granular bed for space-saving in print.}
    \label{fig:impactTypes}
\end{figure}

An impact is simulated by accelerating the projectile, shown in starting position about 2.2cm above the bed in Figure \ref{fig:impactTypes}a, with a one millisecond impulse to achieve desired impact velocity and approach angle. The impactor is also subjected to Earth-like gravity. It is assumed that the impact angle and velocity is minimally impacted by gravity acting on the projectile; for the lowest velocity and shallowest impacts, which are most affected, gravity was observed to change impact angle by only a couple of degrees and velocity by a small fraction of a meter-per-second.

An example snap-shot of each impact behavior type is shown in Figure \ref{fig:impactTypes}. A ricochet event occurs when the impactor departs the initial crater, breaking contact with the granular bed (see Figure \ref{fig:impactTypes}b). The roll-out event occurs when the impactor rolls out of the initial impact crater but does not break contact with the granular surface (see Figure \ref{fig:impactTypes}c). The full-stop impact behavior occurs when the impactor comes to a complete stop in the initial impact crater (see Figure \ref{fig:impactTypes}d).

We carry out two batches of impact simulations. Section \ref{sec:behaviorMapping} reviews the macro behavior of two-hundred eighty six impact simulations with velocities ranging from 1 m/s to 7 m/s and angles between 20\degree \:and 70\degree \:(measured from horizontal). These ranges are selected to be similar to those studied by \cite{Wright2020, Wright2021} to ensure a full span of impactor behaviors. Section \ref{sec:subSurface} studies eighteen simulations; nine at an impact angle of 30\degree \:ranging from 2 m/s to 10 m/s and nine at 60\degree \:over the same velocity range. The majority of impact simulations are conducted across twelve cores to maintain a roughly one-hour wall time. A subset of simulations were run on four, six, and twelve cores and each model returned identical results, confirming that the number of cores does not introduce artefacts to the impact model. To present this work on the same plane as \cite{Wright2020}, we also report the dimensionless Froude number of each impact, calculated in terms of projectile velocity $v$, radius $r$, and gravitational acceleration $g$ by the expression $v/\sqrt{g r}$. 

\section{Results \& Discussions}
\label{sec:results}

\subsection{Impact behavior mapping}
\label{sec:behaviorMapping}

\begin{figure}
    \centering
    \begin{subfigure}[b]{0.49\linewidth}
        \includegraphics[width=3.1in, trim={1cm 0 0 0}, clip]{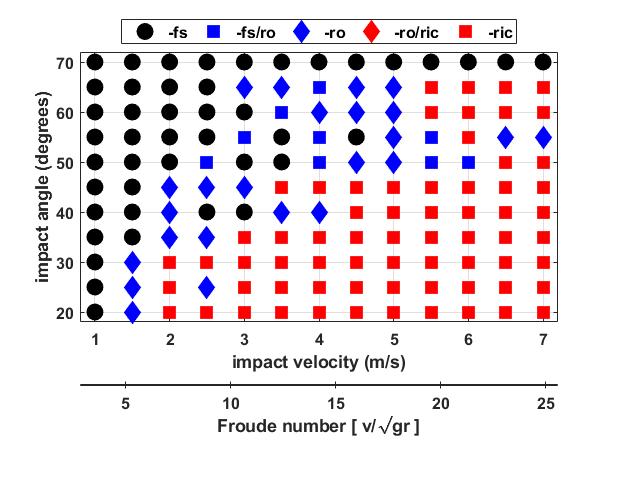}
        \subcaption{ }
    \end{subfigure}
    \begin{subfigure}[b]{0.49\linewidth}
        \includegraphics[width=3.1in, trim={1cm 0 0 0}, clip]{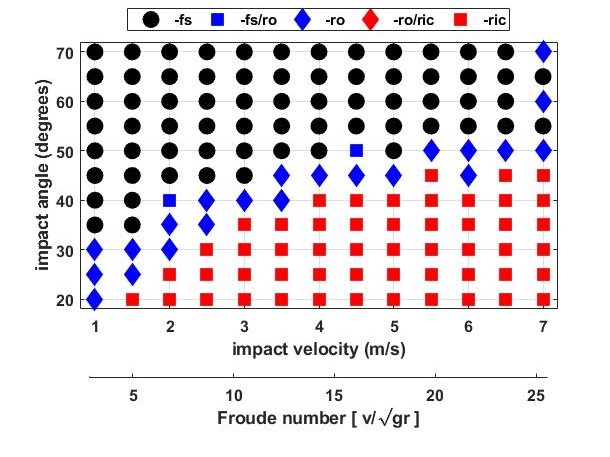}
        \subcaption{ }
    \end{subfigure}
    \caption{Impact event classification as a function of impact angle and impact velocity for (a) granular bed AA from Figure \ref{fig:impactTypes}a and (b) granular bed BB.}
    \label{fig:impactBehaviors}
\end{figure}

In this section, we carry out analysis of the simulations in order to understand how impactor behavior evolves with impact angle and velocity. Denoting each set of initial conditions and their corresponding post-impact behavior as a distinct marker, Figure \ref{fig:impactBehaviors}a shows the impact behavior map obtained from the setup in Figure \ref{fig:impactTypes}a. Red squares represent ricochet impact events (ric), blue diamonds represent roll-out events (ro), and black circles represent full-stop events (fs). Red diamonds represent a hybrid ricochet/roll-out event (ro/ric), where the impactor may have become partially airborne after contact as it moves away from the impact crater. Blue squares represent a hybrid full-stop/roll-out event (fs/ro), where the impact clearly does not settle in the basin of the initial crater but does not completely escape it either.

Inspection of Figure \ref{fig:impactBehaviors}a reveals several features of interest. In the lower right region at higher velocities and shallower impact angles, the ricochet impact response dominates. In the upper left region at lower velocities and steeper impact angles, full-stop behavior dominates. This behavior also appears in a band across the top of the figure, for the steepest impact angle of 70\degree. In-between the full-stop zones and ricochet zones, the roll-out events appear as a transitional behavior. These regions and their relative positions are consistent with experimental findings by \cite{Wright2020}, though it is noted that quantitative agreement is neither expected nor observed due to numerous differences in system specifications such as grain-to-projectile size ratio and the triangular grain geometry. Experimental work is also three dimensional whereas simulations are constrained to two dimensions, which may greatly impact comparisons.

In Figure \ref{fig:impactBehaviors}a, out-of-place roll-out behaviors are noticed at the steepest impact angles and highest velocities. This observation spurs the question, how stable is this behavior map? To answer, we re-generated the granular bed of Figure \ref{fig:impactTypes}a using identical parameters and initial patterning. This produces a `new' granular bed, denoted bed BB (and renaming the original granular system bed AA), that has the same granular properties with marginally different packing characteristics. We repeated the same slew of impact tests on the newly generated bed and classified the impact behaviors, plotting the results in Figure \ref{fig:impactBehaviors}b.

There are clear differences between the plots of Figure \ref{fig:impactBehaviors}. The stochastic roll-out boundary region observed in Figure \ref{fig:impactBehaviors}a is no longer, replaced by a much more regular roll-out region. The ricochet dominated region in Figure \ref{fig:impactBehaviors}b is reduced, with ricochets no longer presenting at steeper impact angles. Nonetheless, there are similarities between the two figures that should not go amiss. Full-stop behaviors still dominate in the upper left regions while ricochets still dominate in the lower right. The `y-intercept' of the boundary between full-stop and ricochet is still close to 20\degree.

\cite{Wright2021} suggest the variation between the two plots is expected due to the small impactor-to-grain size ratio (5.38). They suggest that stochasticity in impact behaviors is anticipated for lower ratios whereas stable and repeatable impact behaviors are produced for higher ratios. Additional to their argument, we theorize that the angular nature of the grains is a factor, exacerbating the effect that the new packing configuration has on the projectile's motion; looser packings allow easier deformation and regular behaviors whereas tighter packings are more interlocked and firm, producing stochastic responses. Tighter packings may also emphasize effects of surface topography which is highly variable due to the angular grain geometry.

Regardless, revealing the systemic factors that cause these variations cannot be fully assessed by visual observation of the impacts alone. In the next section, an algorithm for evaluating packing fraction is presented and the variations between the plots of Figure \ref{fig:impactBehaviors} are re-visited in a quantifiable manner.

\subsection{Sub-surface quantification}
\label{sec:subSurface}

\subsubsection{Packing fraction analysis}
\label{sec:PFstudy}

In this section, post-processing of simulation data from impacts angled at 30\degree \:and 60\degree \:reveal information about the sub-surface granular dynamics that occur during a surface impact. This is a meaningful study because such dynamics contribute to the behavior of the impactor due to the dissipation of energy through grains. Such pathways of dissipation are affected by how grains are positioned relative to their neighbors and the incoming impulse from the impactor. For instance, loose packing of grains could lead to compaction and slippage between grains and a gradual dissipation of energy by friction, creating an overall cushioning effect on an impactor. Tighter grain packings may lead to direct transmission of forces along force chains, producing a stouter impactor behavior reminiscent of solid-on-solid impact.

Impactor behaviors and sub-surface granular mechanics can be tied together by considering changes in grain packing density and its distribution. Typically packing analysis is done by evaluating a system's overall change in packing density, though discrete simulation method can provide a better understanding of such change by considering its spatial distribution. 

The void mapping procedure stems originally from Point Voronoi tessellation \citep{Voronoi1908}, a method that can take a random collection of points and divide the surrounding space into convex polygonal regions belonging to each point. Point Voronoi tessellation may also be used for collections of uniform circles \citep{An2021} or lightly adapted with weighting techniques for poly-disperse circles \citep{Morley2020}. 

However, Point Voronoi tessellation is insufficient for mapping void space between more complex geometries because centroids of particles are no longer equidistant from the edges. Therefore, tessellated regions may cut off parts of the parent grain, causing inaccuracies in packing fraction calculation. This limitation inspired the Set Voronoi tessellation method \citep{Schaller2013, Dong2016, Zhang2021}, an evolution of the Point Voronoi procedure for irregular geometries that uses additional discretization and unification steps. In this work, MATLAB's built-in Point Voronoi function (simply named `voronoi') has been merged with the additional steps of Set Voronoi tessellation to permit accurate division of void space among the triangular grains. The details of the process for two dimensions are described below. With some modification, such a process should be achievable in three-dimensions, but this is beyond the scope of the present paper.

\begin{enumerate}
    \item Raw positional data is read in and rearranged into a more accessible format.
    \item Each grain is shrunk to improve the accuracy of the Voronoi tessellation procedure. Relative to the centroid of each grain, vertex coordinates are adjusted to reduce the size of the grain by a certain factor.
    \item The edges of grains are discretized. Based on a certain length increment, additional points are created between the vertices and added to the overall point-set for each grain.
    \item A set of points bounding the full granular bed is generated to constrain the Voronoi tessellation and maintain accuracy at the edges of the granular bed.
    \item Combining the data points belonging to the grains with the points belonging to the boundary, the whole structure is passed to MATLAB's built-in `voronoi' function which completes the first step of the tessellation.
    \item Resulting polygonal cells created for each point are united together according to their parent grains. The merger of these polygons produces the final `spatial cell' of the granular bed that a particular grain owns.
    \item Finally, a local packing fraction ($\phi_{local}$) is estimated for each grain by dividing the area of the grain itself ($A_{grain}$) by the area of its spatial cell ($A_{cell}$) such that $\phi_{local} = A_{grain}/A_{cell}$.
\end{enumerate}

\begin{figure}
    \centering
    \begin{subfigure}[b]{0.24\linewidth}
        \includegraphics[width=1.15in]{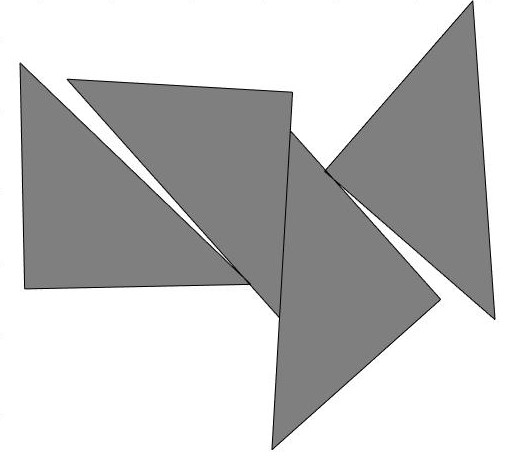}
        \subcaption{Original grain data}
    \end{subfigure}
    \begin{subfigure}[b]{0.24\linewidth}
        \includegraphics[width=1.15in]{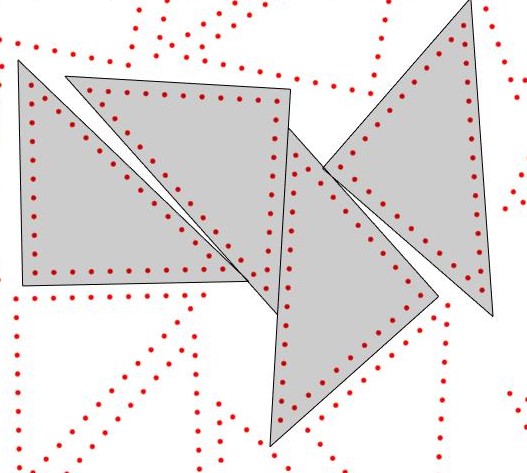}
        \subcaption{Discretized edges}
    \end{subfigure}
    \begin{subfigure}[b]{0.24\linewidth}
        \includegraphics[width=1.15in]{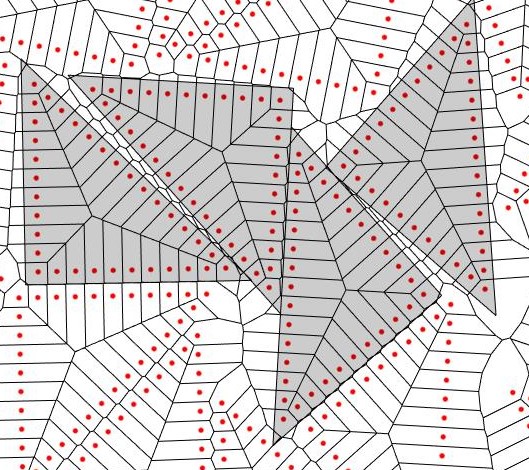}
        \subcaption{Point Voronoi tessellation}
    \end{subfigure}
    \begin{subfigure}[b]{0.24\linewidth}
        \includegraphics[width=1.15in]{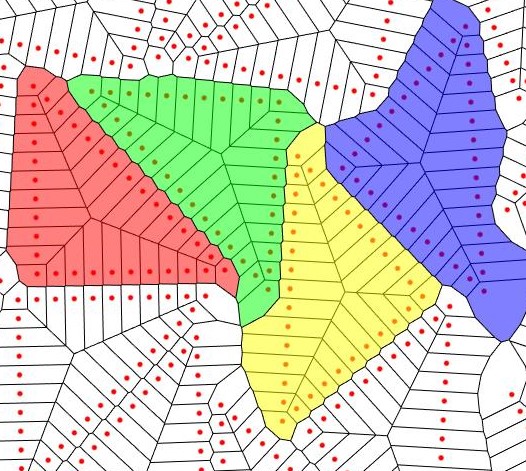}
        \subcaption{Merged final cells}
    \end{subfigure}
    \begin{subfigure}[b]{0.35\linewidth}
        \includegraphics[width=2in,trim={8.5in 2in 7.5in 1.5in}, clip]{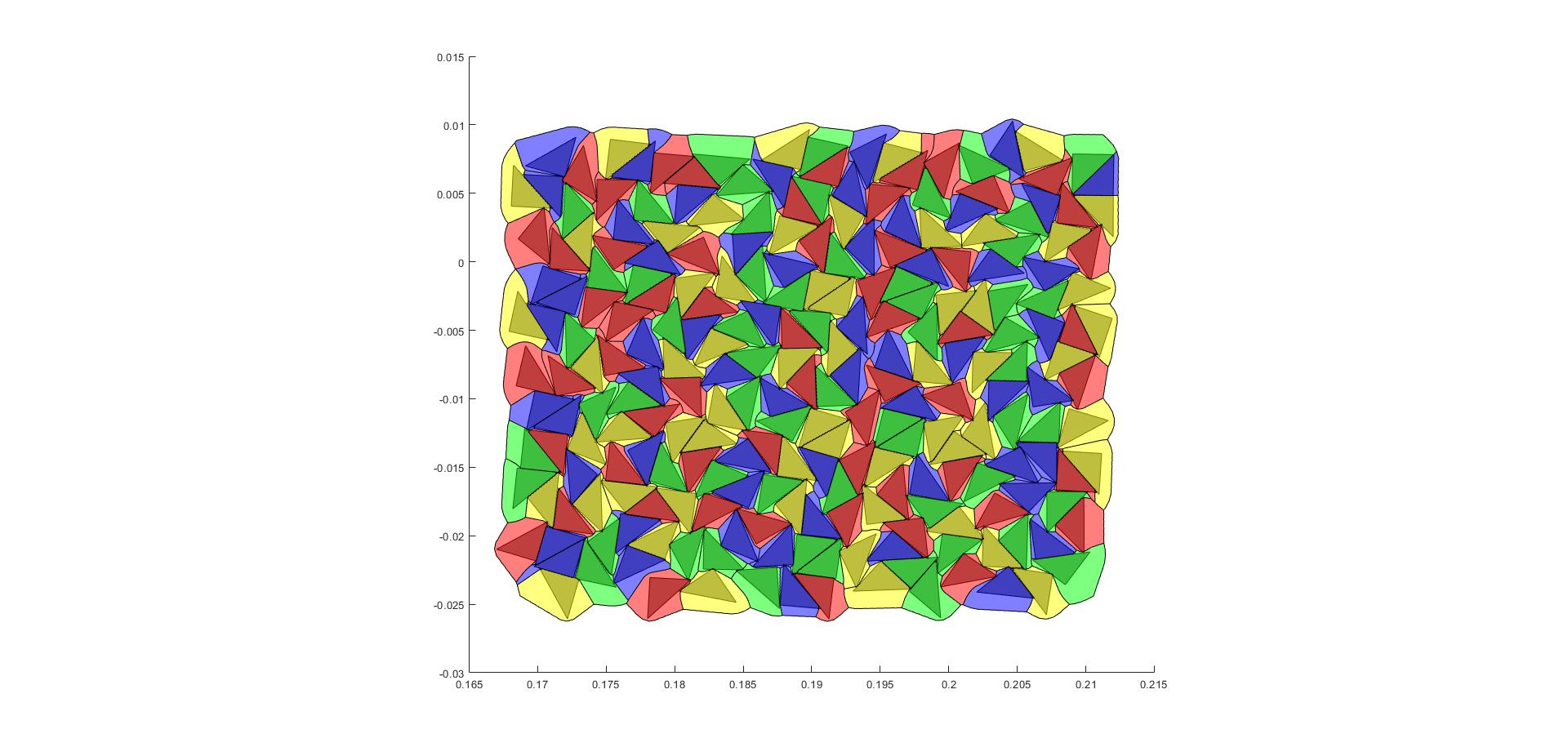}
        \subcaption{Merged final cells}
    \end{subfigure}
    \caption{The tessellation procedure for calculation of grain-level packing fraction.}
    \label{fig:voronoiProc}
\end{figure}

Figure \ref{fig:voronoiProc}a depicts a few grains positioned based on data exported from simulations. Figure \ref{fig:voronoiProc}b depicts these grains after they are shrunken and discretized with additional points along its boundary. Figure \ref{fig:voronoiProc}c shows the result of MATLAB's Voronoi tessellation utility which divides up the space in-between points to spatial allotment closest to a particular point. Figure \ref{fig:voronoiProc}d illustrates how the post-processing script merges the polygons belonging to the larger grain into one larger polygon. Figure \ref{fig:voronoiProc}e shows the final result of the tessellation procedure, with a polygonal cell that represents the region of the granular bed belonging to a grain. The original grains are overlaid in Figure \ref{fig:voronoiProc}e to show the two regions which are used for computing packing fraction. Each color in Figure \ref{fig:voronoiProc}e corresponds to the core that calculated that Voronoi cell, as parallel processing was used to expedite post-processing.

\begin{figure}[h!]
    \centering
    \begin{subfigure}[b]{0.49\linewidth}
        \includegraphics[width=3.0in,trim={0.2in 0.5in 0.6in 0.8in}, clip]{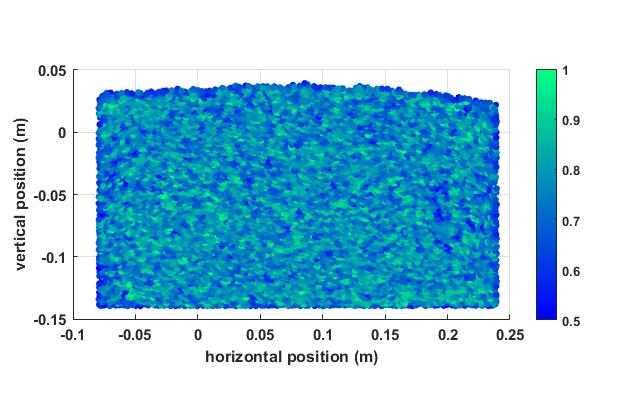}
        \subcaption{}
    \end{subfigure}
    \begin{subfigure}[b]{0.49\linewidth}
        \includegraphics[width=3.0in]{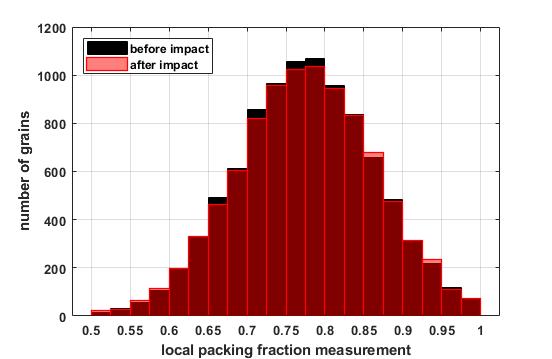}
        \subcaption{}
    \end{subfigure}
    \caption{Evaluation of pre-impact granular bed quality using local packing fraction estimates. A colored packing density plot shows the raw local packing fraction measurements (a), and a histogram (b) illustrates the distribution of these packing fraction measurements.}
    \label{fig:bedQuality}
\end{figure}

The procedure outlined allows for calculation of grain-level packing fraction and such a result is shown in Figure \ref{fig:bedQuality}a, depicting a homogeneous distribution of overall packing density in the granular bed from Figure \ref{fig:impactTypes}a before an impact. A histogram of packing fraction measurements from Figure \ref{fig:bedQuality}a is shown in the black bars of Figure \ref{fig:bedQuality}b. A neat bell curve is produced, indicating a well-settled granular bed. 

\begin{figure}[h!]
    \centering
    \includegraphics[width=4.0in]{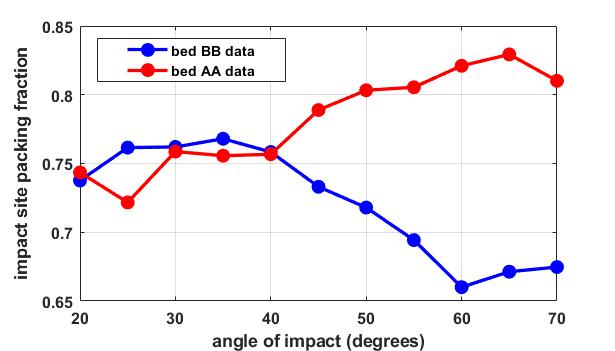}
    \caption{Pre-impact, mean packing fraction of impact site versus impact angle.}
    \label{fig:impSiteMeanPacking}
\end{figure}

Our Voronoi-based packing analysis can re-visit the impact behavior maps of Figure \ref{fig:impactBehaviors} and explain the differences between the two plots.  Figure \ref{fig:bedQuality}a depicts packing measurements for the bed AA associated with Figure \ref{fig:impactBehaviors}a. A similar collection of packing measurements is also collected for bed BB, from Figure \ref{fig:impactBehaviors}b, though not pictured. For each impact angle, the approximate point of impact on the granular bed is determined and the impact site defined as the area of grains within one projectile radius. Using the packing fraction measurements for grains within each impact region, the mean packing fraction of the impact site is calculated. Figure \ref{fig:impSiteMeanPacking} presents this data for both granular beds.

For the data of bed AA (red line in Figure \ref{fig:impSiteMeanPacking}), the mean packing fraction at impact sites is shown to increase at steeper impact angles, reaching a maximum tight packing of 0.83. This supports the stochastic behavior observed at the upper right side in Figure \ref{fig:impactBehaviors}a since it implies that the projectile is hitting a much firmer surface. Therefore, the possibility of a full stop is lessened and there is a higher tendency to display ricochet or roll-out events. In contrast, bed BB data (blue line) displays looser packing at the steeper impact sites, dropping as low as 0.66 locally. This indicates the projectile is hitting a softer granular media that allows greater intrusion, explaining the observation of more full-stop behavior in the upper right of the map in Figure \ref{fig:impactBehaviors}b. Finally, the greatest similarity between the plots of Figure \ref{fig:impactBehaviors} is observed at the shallower impact angles.  Figure \ref{fig:impSiteMeanPacking} also upholds this, since the mean site packing fraction is found to be fairly consistent ($\approx$0.74 - 0.76) between both versions of the granular bed below 45 \degree.

The mechanism causing firmer packings to produce stochastic behaviors is not definitively known. For a hard and flat surface, instead of a granular surface, one would still expect repeatable behavior from an angled impact. We suggest that the reason stochasticity arises in our model is because of the way large grains `rough-up' the surface topography. As the projectile hits a range of different points on the surface, it may encounter a wide range of angled flat faces or even corners. While this condition is also true for looser packings, the firmer packings' diminished ability to dissipate energy through grain sliding exacerbates the topographic effect, producing the stochasticity we observe.

We can also use the packing fraction analysis to assess how the granular bed is re-configured globally and locally by the surface impact. Originally, to demonstrate the reconfiguration, plain packing fraction measurements after impact were assessed in the same manner shown in Figure \ref{fig:bedQuality}a. In Figure \ref{fig:bedQuality}b, the after-impact histogram for a 10 m/s, 30\degree\:impact scenario is overlaid on the pre-impact data with the red bars. The shift is slight and uninformative from this perspective. The poor visualization also suggests that holistic packing fraction for the entire bed, as would be typically measured with fluid displacement or bulk density, would also fail to capture any grain restructuring.

\begin{figure}
    \centering
    \begin{subfigure}[b]{0.31\linewidth}
        \includegraphics[width=2.0in, trim={1cm 2cm 1cm 2cm}, clip]{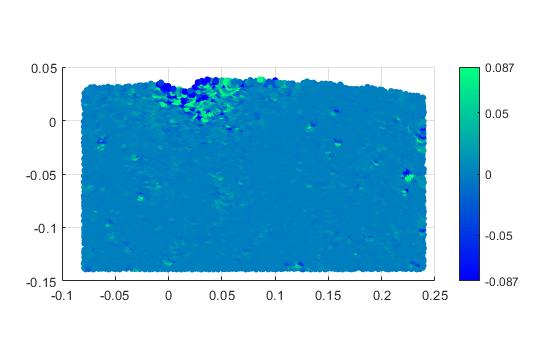}
        \subcaption{v = 2 m/s, Fr = 5.0}
    \end{subfigure}
    \centering
        \begin{subfigure}[b]{0.31\linewidth}
        \includegraphics[width=2.0in, trim={1cm 2cm 1cm 2cm}, clip]{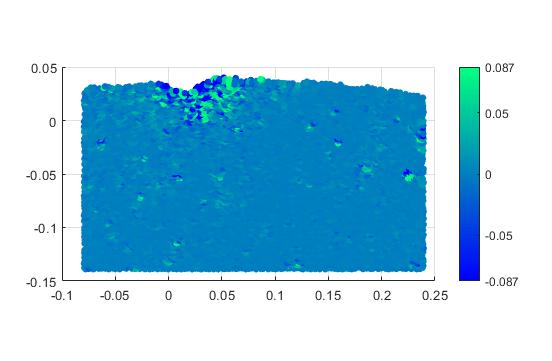}
        \subcaption{v = 3 m/s, Fr = 7.5}
    \end{subfigure}
    \centering
    \begin{subfigure}[b]{0.31\linewidth}
        \includegraphics[width=2.0in, trim={1cm 2cm 1cm 2cm}, clip]{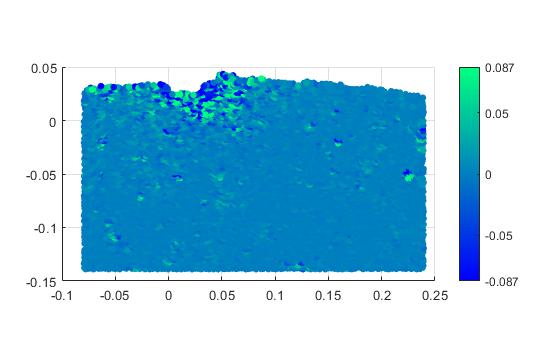}
        \subcaption{v = 4 m/s, Fr = 9.9}
    \end{subfigure}
    \centering
    \begin{subfigure}[b]{0.31\linewidth}
        \includegraphics[width=2.0in, trim={1cm 2cm 1cm 2cm}, clip]{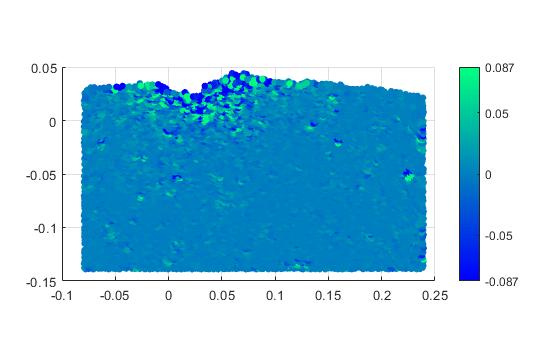}
        \subcaption{v = 5 m/s, Fr = 12.4}
    \end{subfigure}
    \centering
    \begin{subfigure}[b]{0.31\linewidth}
        \includegraphics[width=2.0in, trim={1cm 2cm 1cm 2cm}, clip]{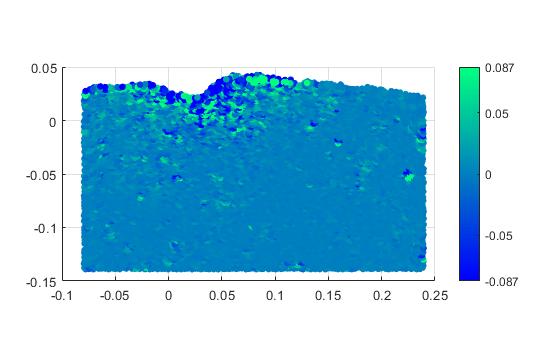}
        \subcaption{v = 6 m/s, Fr = 14.9}
    \end{subfigure}
    \centering
    \begin{subfigure}[b]{0.31\linewidth}
        \includegraphics[width=2.0in, trim={1cm 2cm 1cm 2cm}, clip]{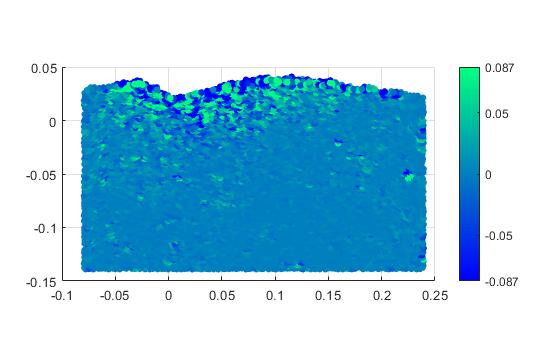}
        \subcaption{v = 7 m/s, Fr = 17.4}
    \end{subfigure}
    \centering
    \begin{subfigure}[b]{0.31\linewidth}
        \includegraphics[width=2.0in, trim={1cm 2cm 1cm 2cm}, clip]{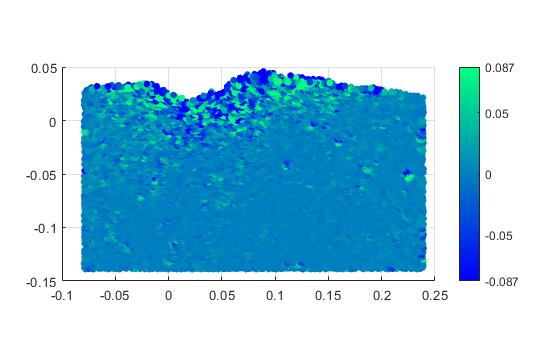}
        \subcaption{v = 8 m/s, Fr = 19.9}
    \end{subfigure}
    \centering
    \begin{subfigure}[b]{0.31\linewidth}
        \includegraphics[width=2.0in, trim={1cm 2cm 1cm 2cm}, clip]{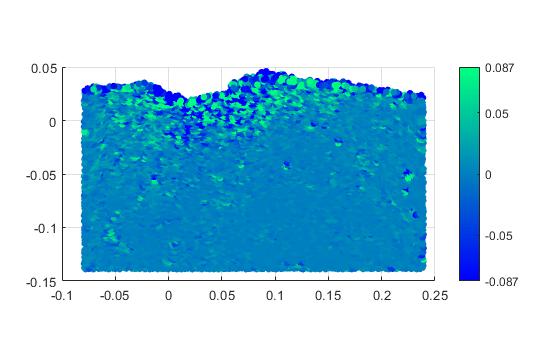}
        \subcaption{v = 9 m/s, Fr = 22.4}
    \end{subfigure}
    \centering
    \begin{subfigure}[b]{0.31\linewidth}
        \includegraphics[width=2.0in, trim={1cm 2cm 1cm 2cm}, clip]{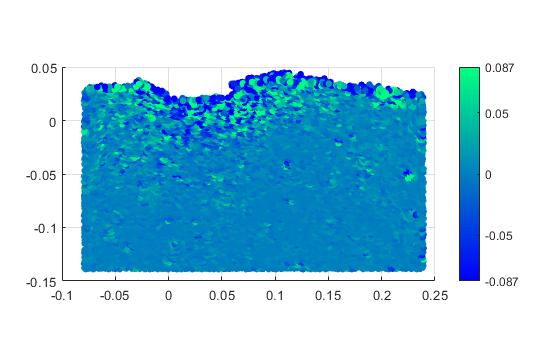}
        \subcaption{v = 10 m/s, Fr = 24.9}
    \end{subfigure}
    \caption{Colored relative packing density plots illustrating the local packing fraction changes after impacts of various velocities at 60\degree \:from the horizontal. Velocity and Froude number of each impact are reported.}
    \label{fig:PFscatters}
\end{figure}

\begin{figure}[h!]
    \centering
    \begin{subfigure}[b]{0.31\linewidth}
        \includegraphics[width=2.0in, trim={1cm 4cm 1cm 2cm}, clip]{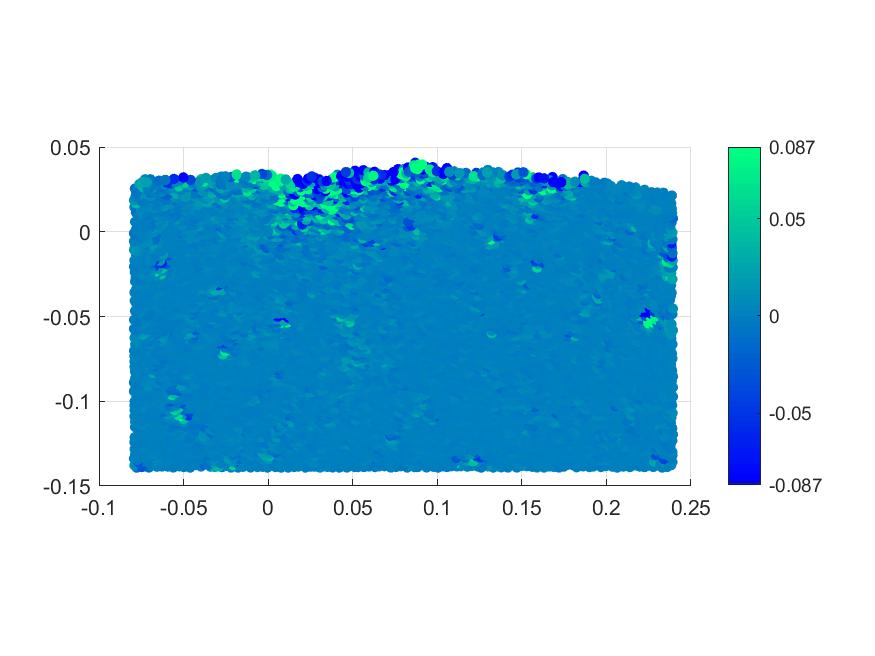}
        \subcaption{v = 2 m/s, Fr = 5.0}
    \end{subfigure}
    \centering
        \begin{subfigure}[b]{0.31\linewidth}
        \includegraphics[width=2.0in, trim={1cm 4cm 1cm 2cm}, clip]{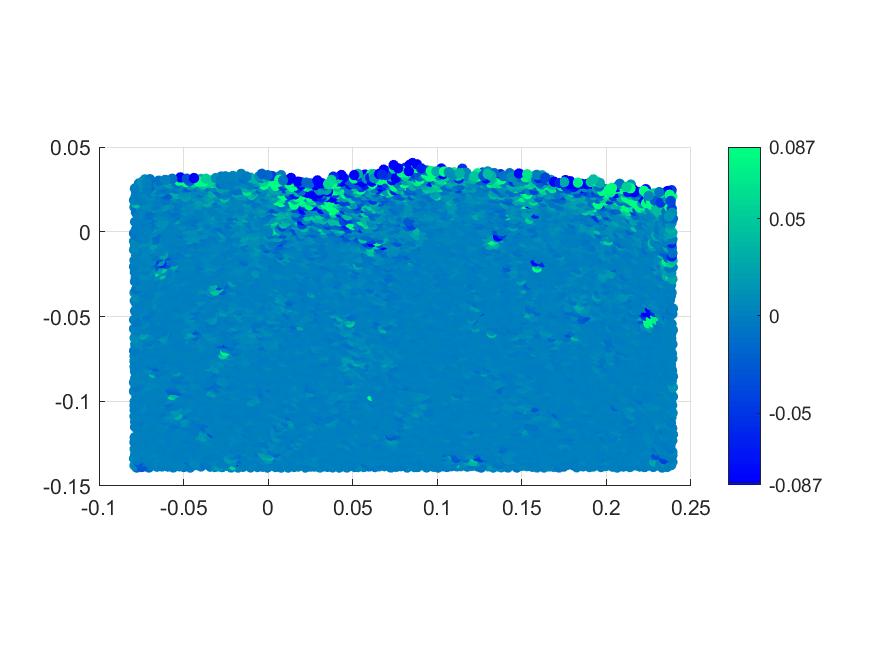}
        \subcaption{v = 3 m/s, Fr = 7.5}
    \end{subfigure}
    \centering
    \begin{subfigure}[b]{0.31\linewidth}
        \includegraphics[width=2.0in, trim={1cm 4cm 1cm 2cm}, clip]{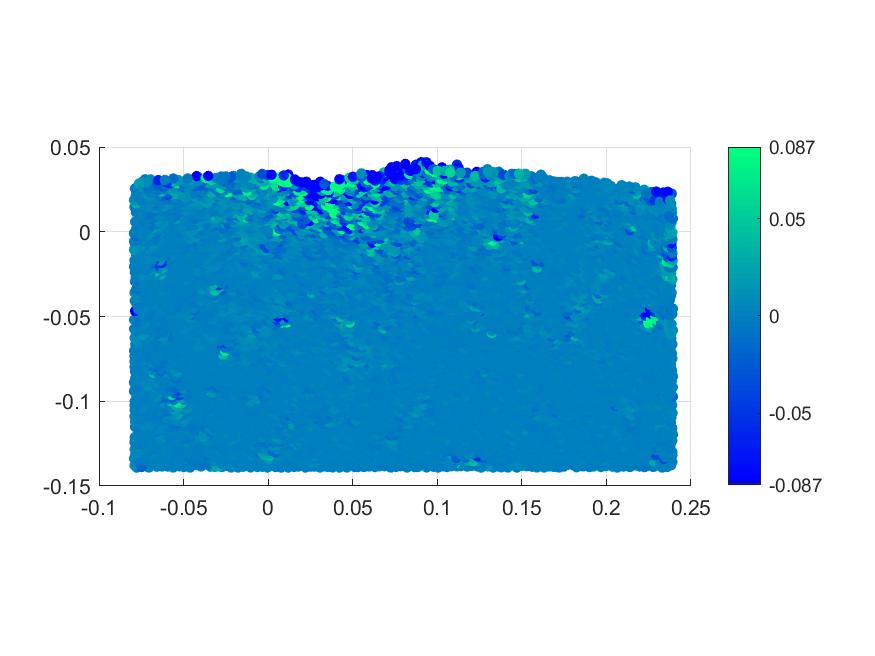}
        \subcaption{v = 4 m/s, Fr = 9.9}
    \end{subfigure}
    \centering
    \begin{subfigure}[b]{0.31\linewidth}
        \includegraphics[width=2.0in, trim={1cm 2cm 1cm 2cm}, clip]{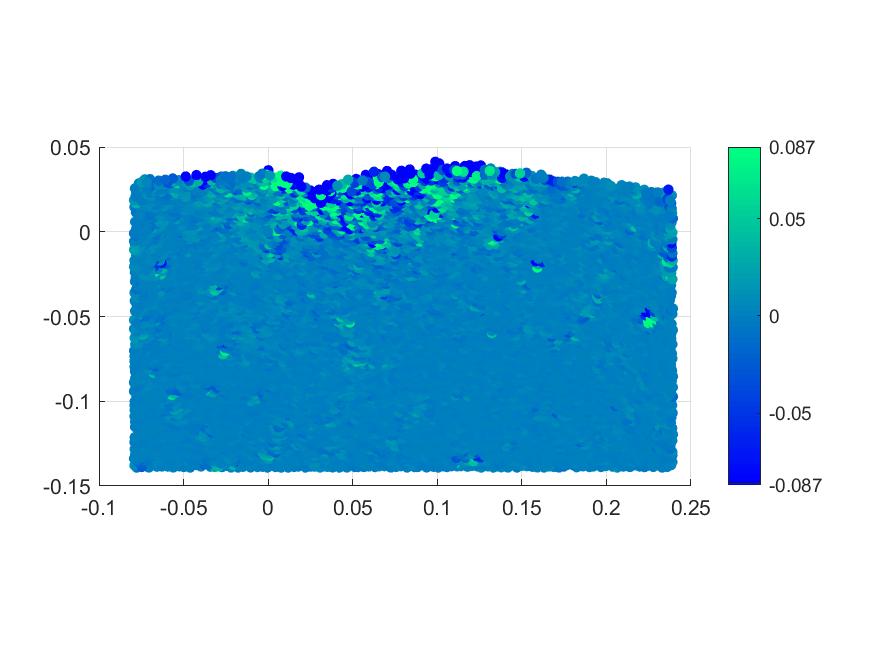}
        \subcaption{v = 5 m/s, Fr = 12.4}
    \end{subfigure}
    \centering
    \begin{subfigure}[b]{0.31\linewidth}
        \includegraphics[width=2.0in, trim={1cm 2cm 1cm 2cm}, clip]{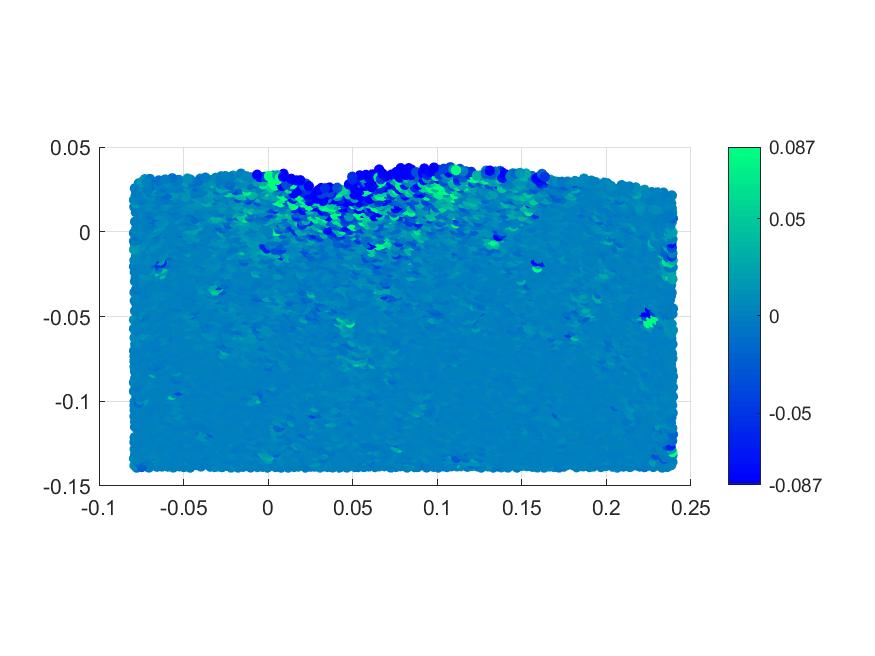}
        \subcaption{v = 6 m/s, Fr = 14.9}
    \end{subfigure}
    \centering
    \begin{subfigure}[b]{0.31\linewidth}
        \includegraphics[width=2.0in, trim={1cm 2cm 1cm 2cm}, clip]{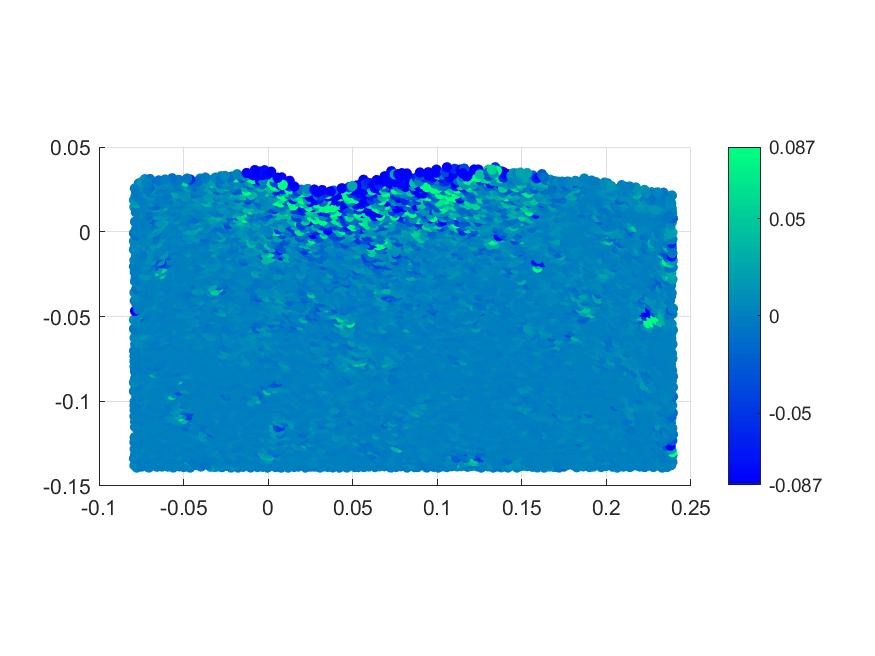}
        \subcaption{v = 7 m/s, Fr = 17.4}
    \end{subfigure}
    \centering
    \begin{subfigure}[b]{0.31\linewidth}
        \includegraphics[width=2.0in, trim={1cm 2cm 1cm 2cm}, clip]{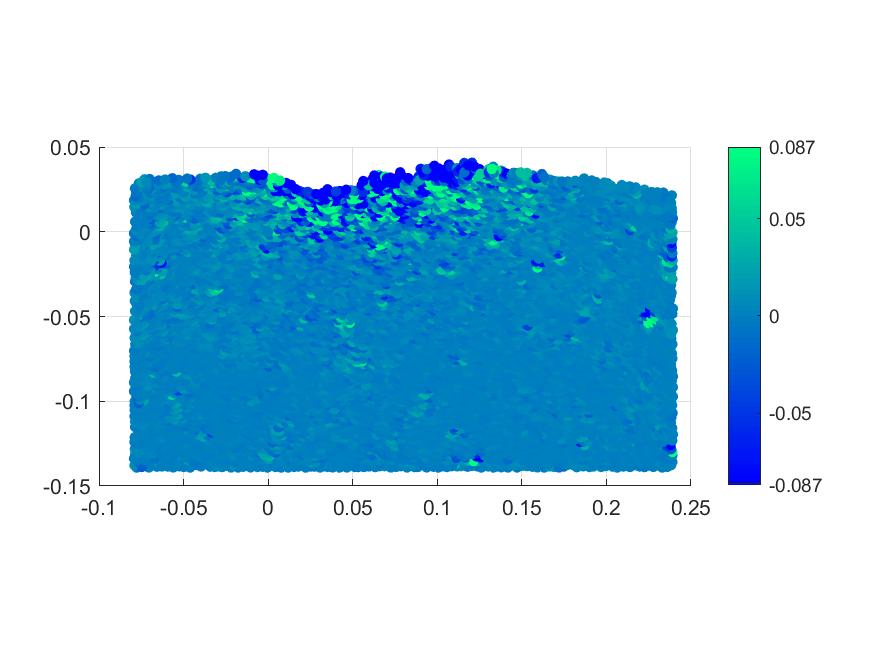}
        \subcaption{v = 8 m/s, Fr = 19.9}
    \end{subfigure}
    \centering
    \begin{subfigure}[b]{0.31\linewidth}
        \includegraphics[width=2.0in, trim={1cm 2cm 1cm 2cm}, clip]{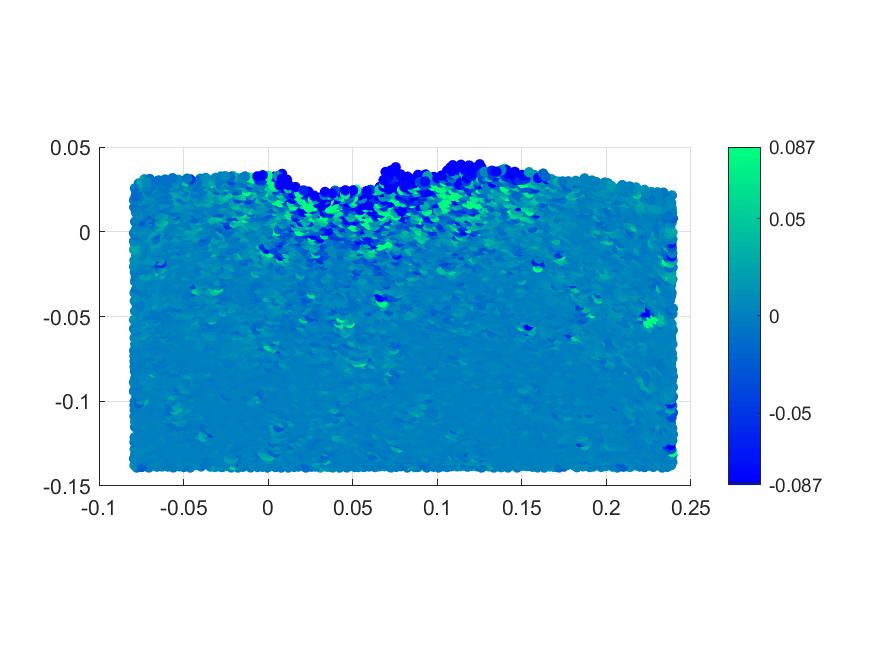}
        \subcaption{v = 9 m/s, Fr = 22.4}
    \end{subfigure}
    \centering
    \begin{subfigure}[b]{0.31\linewidth}
        \includegraphics[width=2.0in, trim={1cm 2cm 1cm 2cm}, clip]{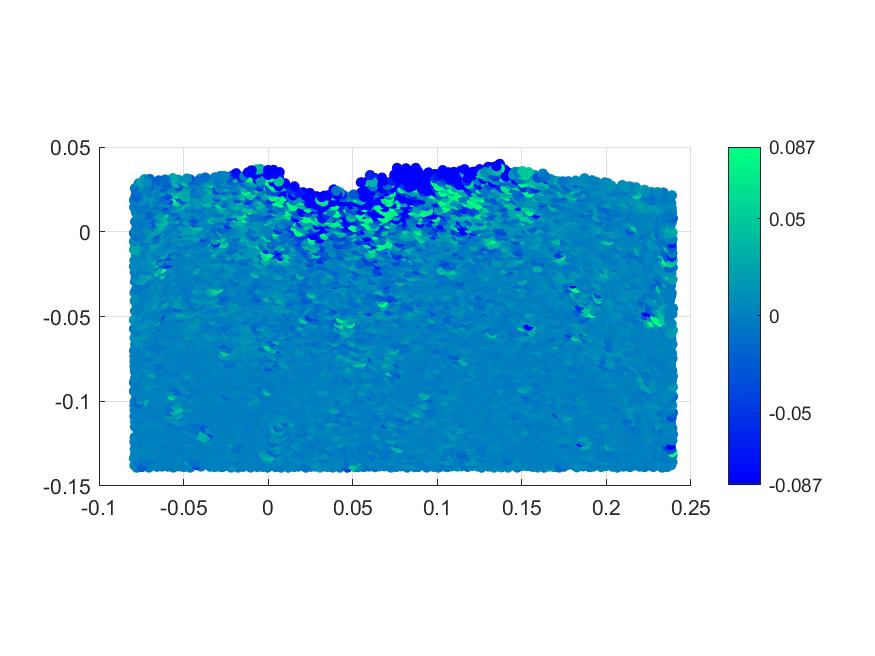}
        \subcaption{v = 10 m/s, Fr = 24.9}
    \end{subfigure}
    \caption{Colored relative packing density plots illustrating results for 30\degree \;impacts.}
    \label{fig:PFscatters30}
\end{figure}

Instead, the per-grain change in packing fraction (relative packing density) was found to be a better metric. Figure \ref{fig:PFscatters} presents packing density plots of each velocity case for 60\degree \:impacts. Results for 30\degree \:simulations can be found in Figure \ref{fig:PFscatters30}. For each grain, the packing fraction post-impact is subtracted from its pre-impact value, so a positive change in packing fraction indicates grain patterns tightening whereas a negative change implies loosening. The upper and lower limits of the colorbar indicate a positive or negative change in packing fraction of approximately 8.7\%, which is taken from the standard deviation of pre-impact packing fraction measurements in Figure \ref{fig:bedQuality}b. The plots of Figure \ref{fig:PFscatters} show how the breadth of disturbance changes with higher velocity impacts.

Considering the 2 m/s density plot (Figure \ref{fig:PFscatters}a), it is shown that only grains immediately around the impact crater reconfigure asymmetrically, within about a one to two impactor diameter radius. Disturbance is contained to this zone, and surrounding surface and sub-surface grains remain relatively undisturbed. 

Figure \ref{fig:PFscatters}e shows that for an impact at 6 m/s the density plot has a region of disturbed grains that has clearly grown with a tripling of the impactor velocity. Interestingly, the right side of the crater shows a zone of grains in which porosity has increased compared to other regions around the crater where grains have packed closer together. Velocities equal to or greater than 6 m/s also show ejecta that has been strewn along the surface, evident from the trail of overall shifted grains extending along the surface beyond the crater.

Finally, at 10 m/s (Figure \ref{fig:PFscatters}i), all aspects of the granular disturbance are amplified. The region of disturbed grains around the crater is roughly about four to five times the diameter of the impactor, the scattered ejecta layer along the surface is much more prominent and several grains deep, and there are even pockets of shifted grains deeper within the granular bed outside of the primary disturbed zone. This latter observation is attributed to structural instabilities within the granular bed that hold up for lower velocities but begin to collapse for higher velocities and indicates that a pressure pulse has propagated through the medium.

These findings intuitively demonstrate how the region of disturbed grains increases steadily with impactor velocity in both the 30\degree \:and 60\degree \:impact cases. The asymmetry of the sub-surface disturbance and ejecta spread is another way in which 30\degree \:and 60\degree \:results may be compared. At the highest velocities of 8 to 10 m/s, the immediate granular disruption underneath the crater is more asymmetrical in the 30\degree \:case, though the more pronounced ejecta spread of 60\degree \:overshadows this. Oddly, the same conclusion does not hold at the lowest velocities of 2 to 4 m/s where sub-surface asymmetry actually appears more pronounced for the 60\degree \:cases. For instance, at 2 m/s, the 60\degree \:impact produces a much deeper and clearly more asymmetrical disturbance compared to 30\degree. These observations suggest that the symmetry of the granular disturbance is not solely influenced by the angle of the impact but also affected by the nature of impact. In the full-stop events of 60\degree \:impacts, much more energy is injected into the granular bed, amplifying the asymmetry at high velocities. The 30\degree \:impacts are predominantly ricochets, so while higher velocities yield increasingly glancing impacts, less energy is injected into the bed and less asymmetry is observed.

\subsubsection{Strain distributions}
\label{sec:strainStudy}

Grain reconfiguration is not necessarily accompanied by a change in packing fraction, so other perspectives should be sought out. Therefore, we compute the strain on each grain using displacement fields. A large-strain Eulerian formulation \citep{Lai2010}, shown in Equation \ref{eqn:eulStrn} in summation index notation, shows how we calculate the strain tensor from the displacement fields of the grains $u$.

\begin{equation}
    \epsilon_{ij} = \frac{1}{2}\left(\frac{\partial u_i}{\partial x_j} + \frac{\partial u_j}{\partial x_i}\right) - \frac{1}{2}\frac{\partial u_m}{\partial x_i}\frac{\partial u_m}{\partial x_j}
    \label{eqn:eulStrn}
\end{equation}
Here, it is implied that we sum over index $m$.

To collapse the strain tensor for each grain to a single representative value, the von Mises distortion \footnote{url{https://dianafea.com/manuals/d944/Analys/node405.html}} is calculated as follows

\begin{equation}
    \boldsymbol{\epsilon_{\text{von Mises}}} = \sqrt{\frac{2}{3}\:\:\left(\boldsymbol{\epsilon}':\boldsymbol{\epsilon}'\right)}.
    \label{eqn:vmStrn}
\end{equation}

\noindent where $\boldsymbol{\epsilon}'$ represents the deviatoric strain tensor, with components given by $\epsilon_{ij}' = \epsilon_{ij} - \frac{1}{3} \:\delta_{ij} \:\epsilon_{ij}$ and $\delta_{ij}$ represents the Kronecker delta function.  The `:' symbol represents a sum over both indices, $\epsilon':\epsilon' = \sum_{ij} \epsilon'_{ij}
\epsilon'_{ij}$.

\begin{figure} [h!]
    \centering
    \begin{subfigure}[b]{0.31\linewidth}
        \includegraphics[width=2.0in, trim={1cm 0 1cm 0}, clip]{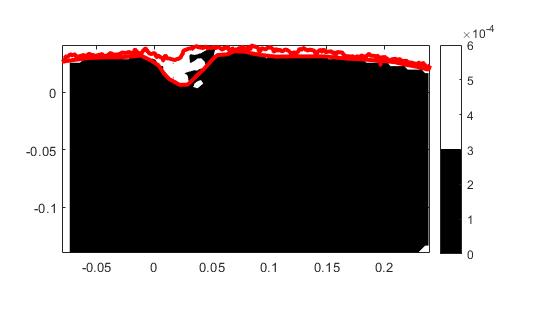}
        \subcaption{v = 2 m/s, Fr = 5.0}
    \end{subfigure}
    \centering
        \begin{subfigure}[b]{0.31\linewidth}
        \includegraphics[width=2.0in, trim={1cm 0 1cm 0}, clip]{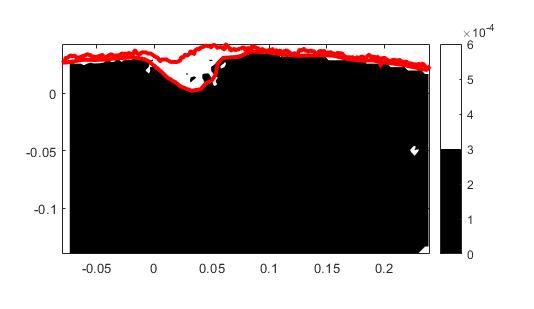}
        \subcaption{v = 3 m/s, Fr = 7.5}
    \end{subfigure}
    \centering
    \begin{subfigure}[b]{0.31\linewidth}
        \includegraphics[width=2.0in, trim={1cm 0 1cm 0}, clip]{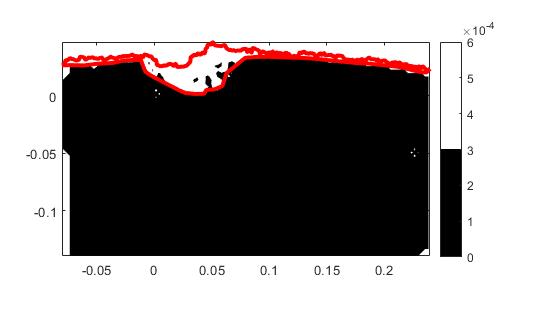}
        \subcaption{v = 4 m/s, Fr = 9.9}
    \end{subfigure}
    \centering
    \begin{subfigure}[b]{0.31\linewidth}
        \includegraphics[width=2.0in, trim={1cm 0 1cm 0}, clip]{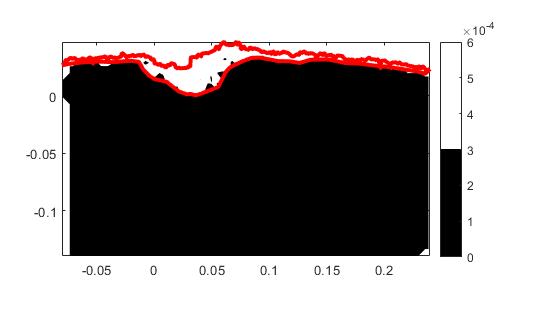}
        \subcaption{v = 5 m/s, Fr = 12.4}
    \end{subfigure}
    \centering
    \begin{subfigure}[b]{0.31\linewidth}
        \includegraphics[width=2.0in, trim={1cm 0 1cm 0}, clip]{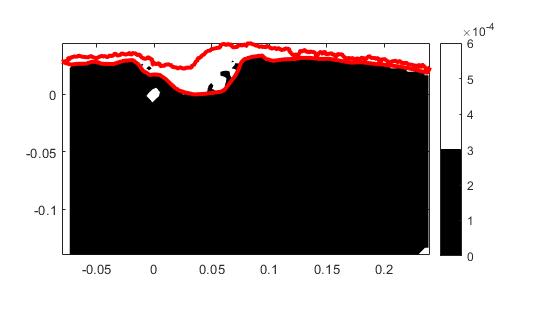}
        \subcaption{v = 6 m/s, Fr = 14.9}
    \end{subfigure}
    \centering
    \begin{subfigure}[b]{0.31\linewidth}
        \includegraphics[width=2.0in, trim={1cm 0 1cm 0}, clip]{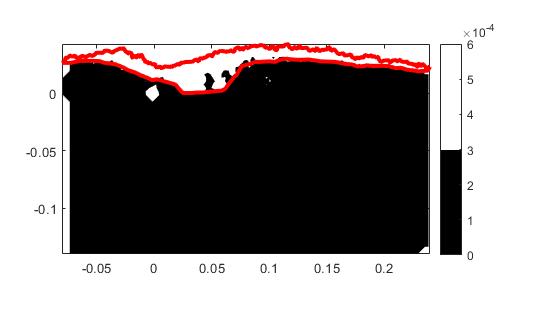}
        \subcaption{v = 7 m/s, Fr = 17.4}
    \end{subfigure}
    \centering
    \begin{subfigure}[b]{0.31\linewidth}
        \includegraphics[width=2.0in, trim={1cm 0 1cm 0}, clip]{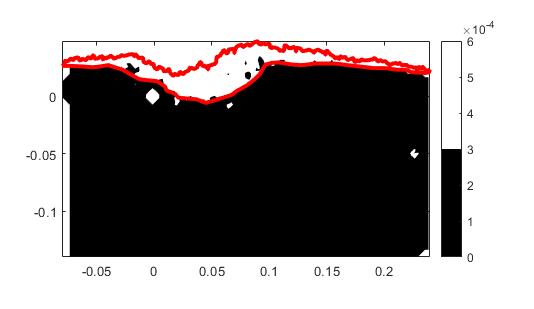}
        \subcaption{v = 8 m/s, Fr = 19.9}
    \end{subfigure}
    \centering
    \begin{subfigure}[b]{0.31\linewidth}
        \includegraphics[width=2.0in, trim={1cm 0 1cm 0}, clip]{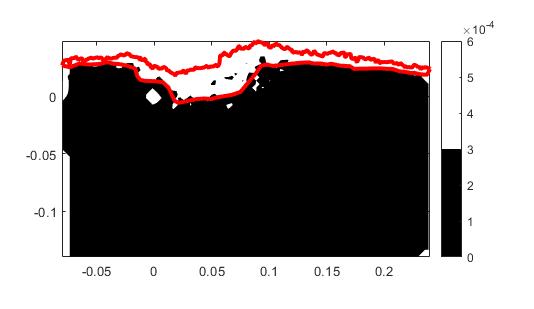}
        \subcaption{v = 9 m/s, Fr = 22.4}
    \end{subfigure}
    \centering
    \begin{subfigure}[b]{0.31\linewidth}
        \includegraphics[width=2.0in, trim={1cm 0 1cm 0}, clip]{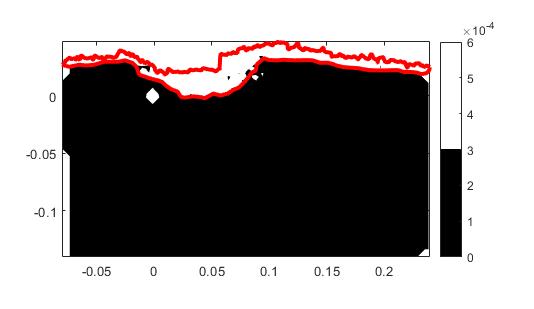}
        \subcaption{v = 10 m/s, Fr = 24.9}
    \end{subfigure}
    \caption{Contour plots of Von Mises strain distribution in the granular bed after impacts of various velocities at 60\degree \:from the horizontal. Velocity and Froude number of each impact are reported. Red lines bound the white strained region of material, denoted the `skin zone'. Black indicates granular media that is not strained beyond the quasi-static flow strain threshold.}
    \label{fig:strnCntrs}
\end{figure}

\begin{figure}[h!]
    \centering
    \begin{subfigure}[b]{0.31\linewidth}
        \includegraphics[width=2.0in, trim={1cm 0 1cm 0}, clip]{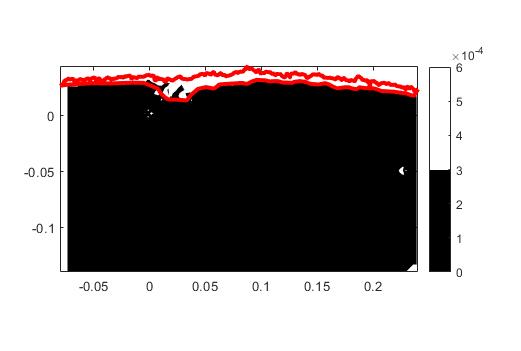}
        \subcaption{v = 2 m/s, Fr =  5.0}
    \end{subfigure}
    \centering
        \begin{subfigure}[b]{0.31\linewidth}
        \includegraphics[width=2.0in, trim={1cm 0 1cm 0}, clip]{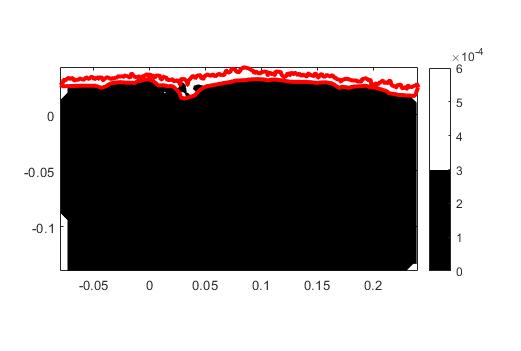}
        \subcaption{v = 3 m/s, Fr = 7.5}
    \end{subfigure}
    \centering
    \begin{subfigure}[b]{0.31\linewidth}
        \includegraphics[width=2.0in, trim={1cm 0 1cm 0}, clip]{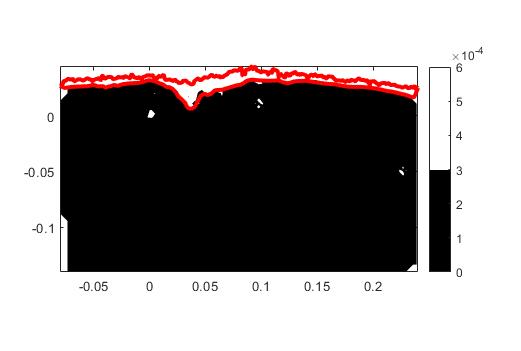}
        \subcaption{v = 4 m/s, Fr = 9.9}
    \end{subfigure}
    \centering
    \begin{subfigure}[b]{0.31\linewidth}
        \includegraphics[width=2.0in, trim={1cm 0 1cm 0}, clip]{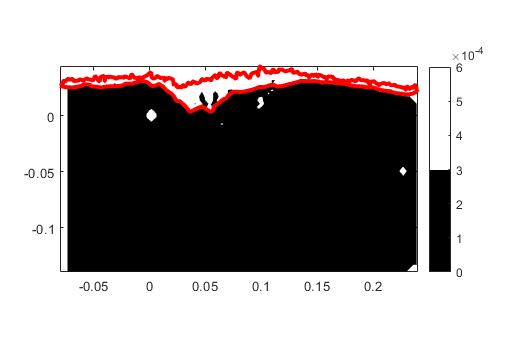}
        \subcaption{v = 5 m/s, Fr = 12.4}
    \end{subfigure}
    \centering
    \begin{subfigure}[b]{0.31\linewidth}
        \includegraphics[width=2.0in, trim={1cm 0 1cm 0}, clip]{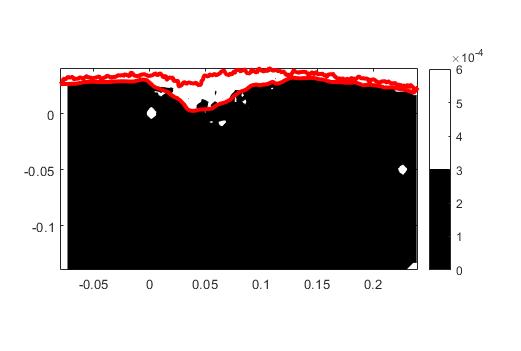}
        \subcaption{v = 6 m/s, Fr = 14.9}
    \end{subfigure}
    \centering
    \begin{subfigure}[b]{0.31\linewidth}
        \includegraphics[width=2.0in, trim={1cm 0 1cm 0}, clip]{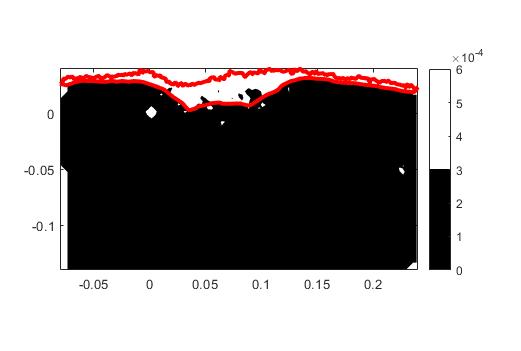}
        \subcaption{v = 7 m/s, Fr = 17.4}
    \end{subfigure}
    \centering
    \begin{subfigure}[b]{0.31\linewidth}
        \includegraphics[width=2.0in, trim={1cm 0 1cm 0}, clip]{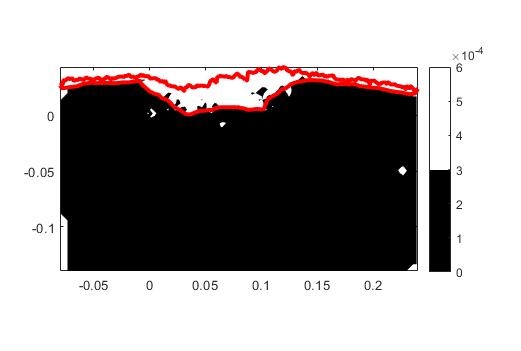}
        \subcaption{v = 8 m/s, Fr = 19.9}
    \end{subfigure}
    \centering
    \begin{subfigure}[b]{0.31\linewidth}
        \includegraphics[width=2.0in, trim={1cm 0 1cm 0}, clip]{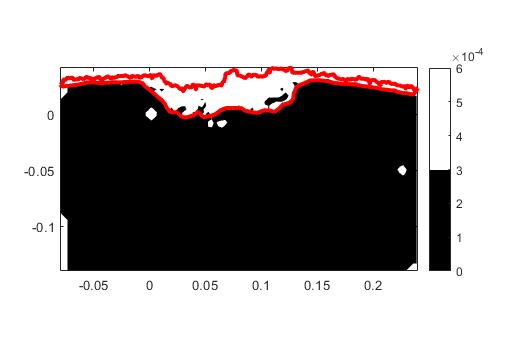}
        \subcaption{v = 9 m/s, Fr = 22.4}
    \end{subfigure}
    \centering
    \begin{subfigure}[b]{0.31\linewidth}
        \includegraphics[width=2.0in, trim={1cm 0 1cm 0}, clip]{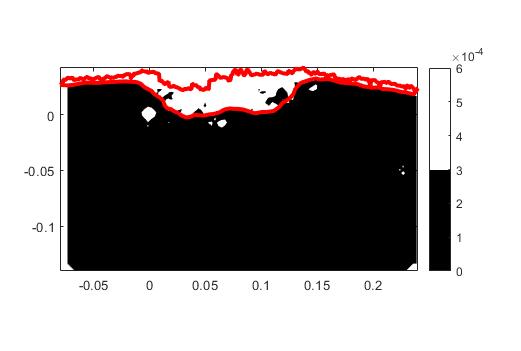}
        \subcaption{v = 10 m/s, Fr = 24.9}
    \end{subfigure}
    \caption{Contour plots of Von Mises strain distribution for 30\degree \:impacts.}
    \label{fig:strnCntrs30}
\end{figure}

To visualize the strained region around the impact crater, a two-color contour plot is created, the threshold of which is set at a strain level for the von Mises distortion $\epsilon_{\text{von Mises}} = 3 \times 10^{-4}$. The colors, then, imply that white region has permanently deformed while the black region has deformed within elastic limit. This value is based on a quasi-static flow strain measurement calculated from separate simulation. Using the same granular bed, a 5 cm horizontal rigid plate is positioned just above the surface. A displacement boundary condition pushes the plate downward at a slow rate of 0.25 mm/s for 1 cm, breaking through the granular surface. Reading the reaction force against the plate reveals the time at which the surface yields and then post-processing the grain displacement data at that frame gives the local strain corresponding to the yield, found to be $3 \times 10^{-4}$. 
Figure \ref{fig:strnCntrs} displays the strain contours for the 60\degree \:impact cases. Figure \ref{fig:strnCntrs30} shows results for 30\degree. Within these plots, the white region bound by red lines is the granular material strained beyond the quasi-static flow strain threshold. The black region is granular media that is not strained beyond that point. The upper red line marks the surface of the granular media, the lower red line follows the contour boundary.

Figure \ref{fig:strnCntrs}a shows the distortion contour plot for the lowest velocity 2 m/s impact. As was shown with the relative packing fraction results, the strained region of grains is largely confined to the impact crater. A moderate velocity level of 6 m/s in Figure \ref{fig:strnCntrs}e demonstrates a larger region of strained grains with a noticeable strained layer to the right of the impact crater, indicating ejecta along the surface. In Figure \ref{fig:strnCntrs}i at 10 m/s, all features of the 6 m/s case are further amplified except, seemingly, for the depth of the perturbed region. Otherwise, the layer of surface-scattered material is much deeper as is the span of the directly strained region around the impact site. Figure \ref{fig:strnCntrs} also illustrates the asymmetry in substrate response caused by the oblique impact mentioned in the packing fraction analysis.

The observation that depth of the strained region did not continue to grow for higher velocity impacts at 60\degree \:spurred further study of the strain contours. The enclosure of the two red lines, marking the strained media, is denoted the `skin zone'. The skin zone, therefore, represents the region of grains that have been permanently strained beneath the surface as well as the layer of ejected grains that have settled atop the surface. The difference in height between corresponding points of upper and lower boundaries produces a `skin depth' measurement.

\begin{figure} [h!]
    \centering
    \begin{subfigure}[b]{0.49\linewidth}
        \includegraphics[width=2.9in, trim={0 0 0 0}, clip]{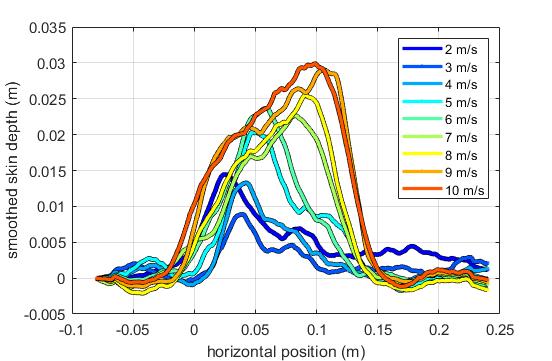}
        \subcaption{30\degree \:impact data}
    \end{subfigure} 
    \centering
    \begin{subfigure}[b]{0.49\linewidth}
        \includegraphics[width=2.9in, trim={0 0 0 0}, clip]{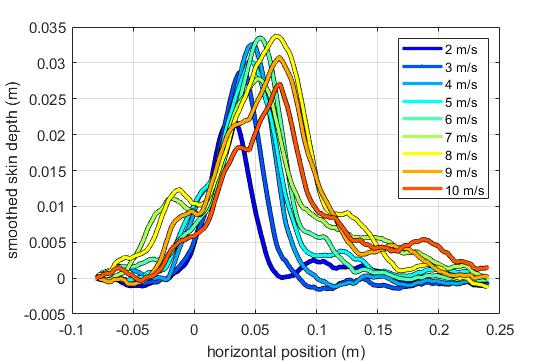}
        \subcaption{60\degree \:impact data}
    \end{subfigure}
    \caption{Plots of skin depth - the difference in height between corresponding points on the upper and lower boundaries shown in versus horizontal position.}
    \label{fig:skinPlots}
\end{figure}

Figure \ref{fig:skinPlots} shows skin depth measurements versus horizontal position for 30\degree \:and 60\degree \:impacts. For easier viewing, the results are smoothed with a moving average and a vertical shift to align the leftmost points. The smoothing helps to view the trend in skin zone geometry with increasing velocity impacts. In Figure \ref{fig:skinPlots}a, showing 30\degree \:impacts, a fairly straightforward trend is seen where the skin zone deepens and widens in a steady manner as velocity increases. Contrast to this the 60\degree \:impacts in Figure \ref{fig:skinPlots}b. Starting from the blue line, representing the 2 m/s case, a single sharp peak is shown with a general return to zero moving further to the right. The next five impact cases, up to about 6 m/s, show that same sharp peak continue to grow accompanied by a levelling out of the zone moving to the right. However, for 7 m/s impacts and above, the peak no longer grows and instead reduces in height and widens out as lateral strain appears to dominate over strains deeper in the bed.

The observed trend of peaks in the skin zone depth measurements for 60\degree \:suggests a potential change in the response of the granular media to the surface impacts of increasing velocity. Up until a point, the strained region does grow deeper with increasing velocity. However, around 7 m/s for this specific scenario, the peak shortens with deeper permanently deformed regions spanning further to the left and right of the impact crater. We have not been able to identify this type of trend in literature though it suggests that the energy dissipation mechanism  that arrests projectile motion may change with increasing velocity, leading to a more lateral plastic deformation.

\begin{figure} [h!]
    \centering
    \includegraphics[width=3.1in, trim={0 0 0 0}, clip]{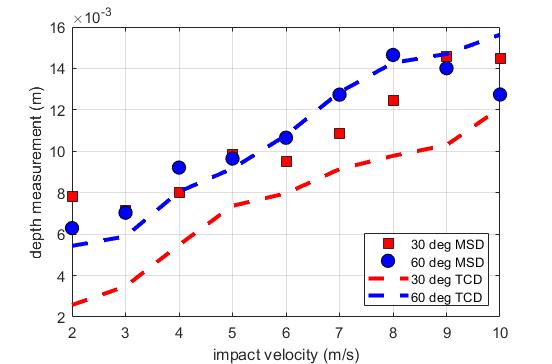}
    \caption{Plots of Mean Skin Depth (MSD) and measured Transient Crater Depth (TCD) for 30\degree \:(red) and 60\degree \:(blue) impacts.}
    \label{fig:meanSkinPlot}
\end{figure}

To further investigate this trend's dependence on velocity, the raw skin depth measurements are averaged to create a data point for mean skin depth at corresponding impact velocity. The data are displayed in Figure \ref{fig:meanSkinPlot}, where mean skin depth (MSD) measurements for 30\degree \:impacts and 60\degree \:impacts are marked by red squares and blue circles, respectively. Additionally, for each simulation, measurements of the transient crater depth (TCD) - the deepest point to exist in the crater during the impact event - are also displayed for 30\degree \:and 60\degree \:impacts as red and blue dashed lines.

From the four data sets of Figure \ref{fig:meanSkinPlot}, some interesting discussions arise. The MSD data from both 30\degree \:and 60\degree \:simulations roughly coincide each other until around 9 m/s, despite predominantly ricochet behavior in the 30\degree \:impact simulations and full-stop behavior in the 60\degree \:cases. For instance, at the 5m/s case, the mean skin depth of the 30\degree \:impact is near identical to 60\degree. The TCD data, on the other hand, show that crater depths are generally greater for across the velocity range for 60\degree \:impacts than for 30\degree \:impacts. Furthermore, the TCD data for 60\degree \:also closely aligns with the MSD data for 60\degree. This last observation suggests that for such an angle in these specific conditions, measuring crater depth at the surface directly reveals the mean depth of the disturbed grain region beneath. The analysis of skin depth reveals that for steeper impacts, the trends of depth of the skin zone is expected to break at smaller velocities since the granular bed can perform as a monolithic solid at larger depths from the free surface.

\section{Conclusion}
\label{sec:conclusion}

We have presented simulations that study low-velocity oblique impacts into a non-spherical granular media. Citing trends in granular dynamics literature toward modeling angular grain geometries as a better representation of non-terrestrial regolith, this work has developed discrete 2D simulations based on finite elements, a technique called the Discrete Finite Element Method (DFEM), that models the simplest possible polygonal grain geometry – a triangle. Maintaining Earth-like gravity to enable comparison to experimental literature, impacts at various angles and velocities of a circular projectile into a bed of triangular grains have been presented. Our results demonstrate the three classes of impact behavior and reproduce the velocity and impact angle dependent division between ricochet and roll-out behavior, recently observed in experiments \citep{Wright2020}.

We find that visualizing the per-grain local packing fraction variance within 8.7\% reveals a sub-surface perturbed region that is lopsided with respect to the impact point.  Local porosity variations are focused in front of the impact site by a distance that is similar in size to the crater radius. The region in front of the impact can be compacted while substrate nearer the surface increases in porosity. By strain analyses, we introduce the notion of `skin zone', or the region beneath the surface that is permanently deformed  beyond the quasi-static yield threshold. Measurements of this skin zone have demonstrated trends that suggest granular dissipative mechanisms change when faced with increasing impact velocity,especially at steeper impact angles. 

There is more work to do. Mono-disperse triangular grains are not reflective of any naturally occurring material and their use here was to ease development of the DFEM model. Various gravity conditions, extending to micro-gravity, will be studied. Polydispersity, another defining characteristic of regolith, will also need to be considered to bring simulations closer in line with the environments we wish to study. The finite elements basis also permits many more perspectives beyond the grain coordinate and displacement based metrics we have evaluated in this paper, such as nodal data. Outputting fields such as nodal stress direct from the impact model will help us to visualize force networks that propagate throughout granular media. An example of this output is shown in Appendix II for a 3.5 m/s impact at 45\degree, as a showcase for future efforts. While detailed analysis of stress contours is beyond the scope of this paper, it will be a prime focus of the next steps in this modeling work.

Developing a comprehensive understanding of the phenomena entailed in low-velocity impacts would support space missions targeting asteroid and planetary exploration in a myriad of ways. Our work demonstrates the importance of packing fraction in determining the behavior of an impactor, an especially key realization for granular media such as regolith with very high packing potential. Through our strain analyses, we have  shown the existence of and measured the depth to which granular media is disrupted, a finding that, when reversed, informs on how deep the composition of a granular bed must be understood to anticipate the behavior of a surface impact. Ultimately, these conclusions could have ripple effects into other areas of research beyond just impact events. Improved simulations of regolith media and better virtualization of its behavior could aid development of surface vehicles, methods for terrain cultivation and terraforming, drilling equipment, or in-situ building materials.

\section*{Acknowledgements}
This material is based on work partially supported by NASA grant 80NSSC21K0143 

\printcredits

\bibliographystyle{cas-model2-names}

\bibliography{references.bib}

\newpage
\section*{Appendix I: Verifying the model performance}

\begin{table}[h!]
\centering
    \caption{Parameter setup for LAMMPS simulation. *\textit{Note: The projectile density is reduced to ensure the same projectile-to-grain mass ratio as for the Abaqus DEM simulation.}}
    \begin{tabular}{|c|c|c|c|c|c|}\hline
     projectile & projectile & grain & grain & Young's & friction \\
     diam. ($m$) & density* ($kg/m^3$) & diam. ($m$) & density ($kg/m^3$) & modulus ($GPa$) & ceoff \\\hline
     0.01615 & 541.7 & 0.002394 & 2600 & 7.0 & 0.84 \\\hline
    \end{tabular}
    \label{tab:LAMMPSsetup}
\end{table}

Finite element method is not often applied to model discrete grains. A primary difference between the triangular grain and a DEM spherical grain is the number of nodes per discrete element that must be managed. Spherical DEM only has one node and a fixed radius, allowing for high optimization of simulation codes and efficient algorithms for contact and friction controls. In contrast, the right triangle grains have three nodes and flat sides allowing for more complex inter-grain contact to arise. Because of this, the contact algorithms employed by Abaqus must be more versatile. This is examined using two parallel simulations. The first simulation is of an impact into a bed of spheres, modeled in Abaqus and thus dependent on Abaqus' contact algorithm. The second simulation tests the same scenario, but with the model built in LAMMPS per the parameters in Table \ref{tab:LAMMPSsetup}. LAMMPS, an open source molecular dynamics software \citep{LAMMPS}, was selected as it is a widely used discrete granular systems solver. 

In both simulations, the impactor connects with the surface at an angle of 60\degree\:from the horizontal and a velocity of 4 m/s.

\begin{figure}[h!]
    \centering
    \begin{subfigure}[b]{0.49\linewidth}
        \includegraphics[width=2.4in, trim={0 0 0 2cm}, clip]{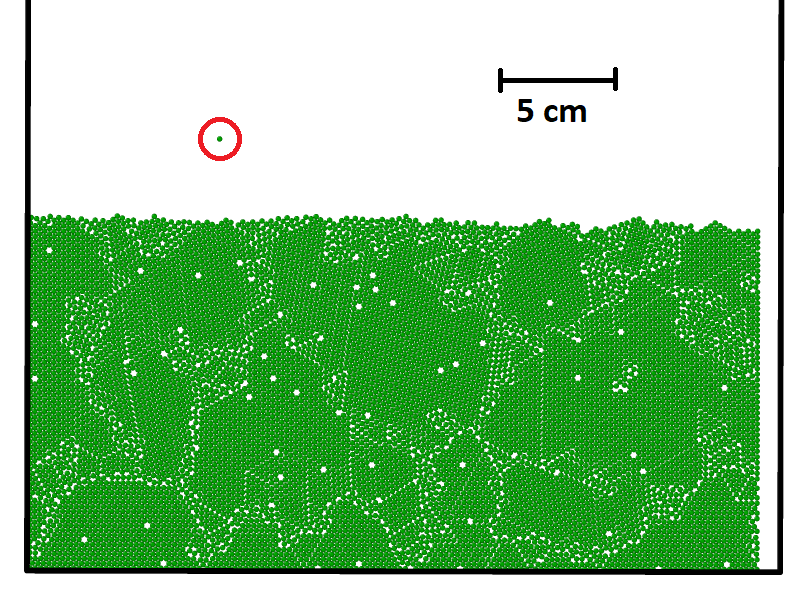}
    \end{subfigure}
    \begin{subfigure}[b]{0.49\linewidth}
        \includegraphics[width=2.4in, trim={0 0 0 2cm}, clip]{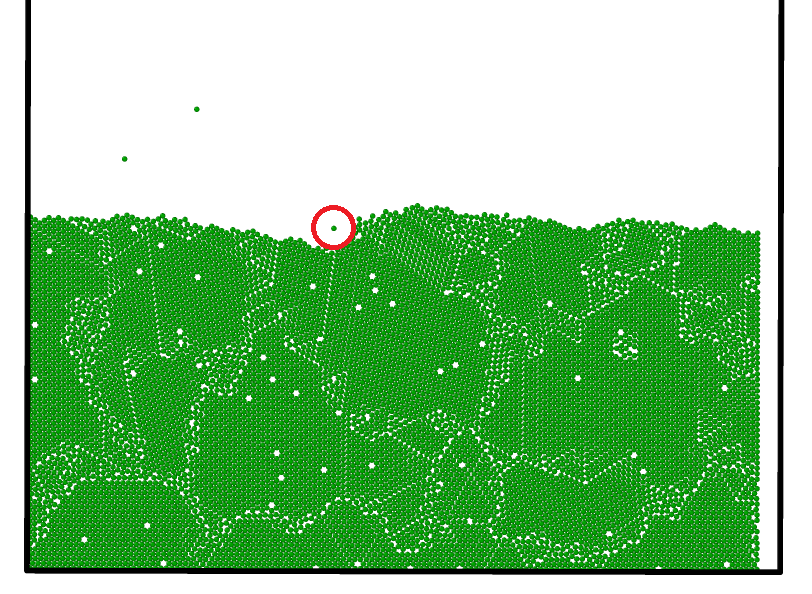}
    \end{subfigure}
    \caption{Initial and final frames of LAMMPS impact simulation. The particle representing the projectile is circled in red.}
    \label{fig:verLAMMPS}
\end{figure}

\begin{figure}[h!]
    \centering
    \begin{subfigure}[b]{0.49\linewidth}
        \includegraphics[width=2.4in, trim={0 0 0 2cm}, clip]{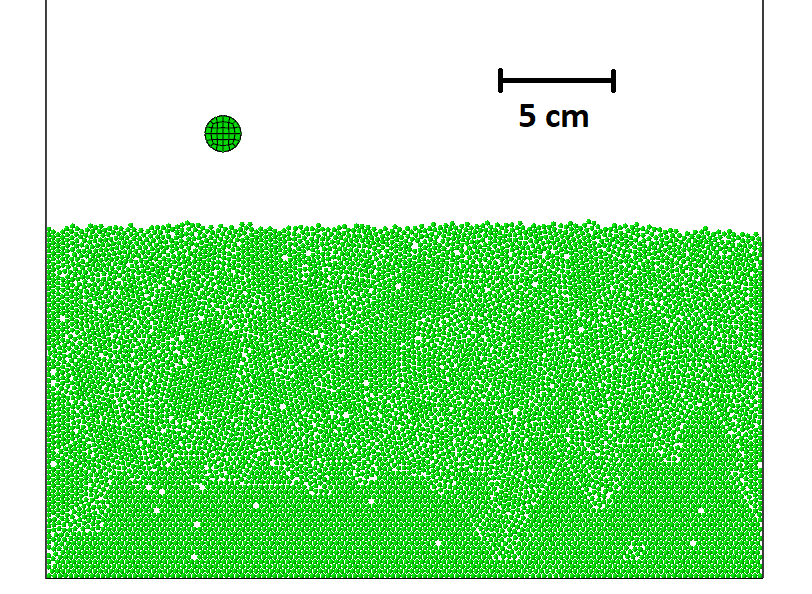}
    \end{subfigure}
    \begin{subfigure}[b]{0.49\linewidth}
        \includegraphics[width=2.4in, trim={0 0 0 2cm}, clip]{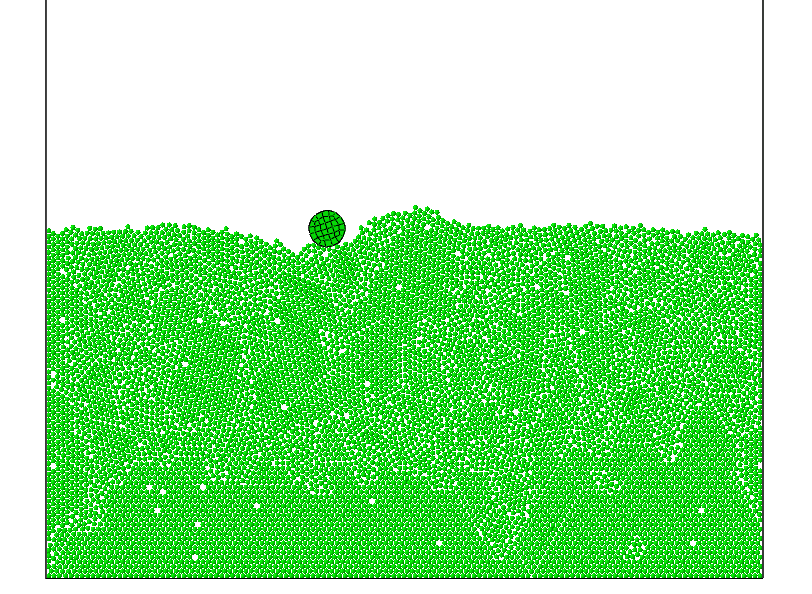}
    \end{subfigure}
    \caption{Initial and final frames of Abaqus impact simulation.}
    \label{fig:verABAQUS}
\end{figure}

Figures \ref{fig:verLAMMPS} and \ref{fig:verABAQUS} depict the initial and final frames of the impact simulations carried out in LAMMPS and Abaqus, respectively. Qualitatively, it is observed that both simulations result in a clear full-stop of the impactor. The width and depth of both craters are similar and the `mounds' of material at the leading edges of the craters are of comparable size. Despite the minute differences in how the LAMMPS and Abaqus models are formulated in terms of material properties and contact controls, it is clear that both platforms performed similarly. 

In terms of computational efficiency, LAMMPS was found to perform faster than both Abaqus DEM and DFEM models. Considering bed-settling and a single impact modeling duration, LAMMPS simulations finished in about 15 minutes on a single core while Abaqus DEM finished in about 5.5 hours on a single core. In addition, the DFEM simulations with three times as many nodes finished in about seven hours on 12 cores. However, Abaqus allows a wider range of tailoring in model setup and simulation parameters compared to LAMMPS, which justifies its use despite slower performance.

\begin{figure}[h!]
    \centering
    \begin{subfigure}[b]{0.49\linewidth}
        \includegraphics[width=2.4in, trim={0 0 0 0}, clip]{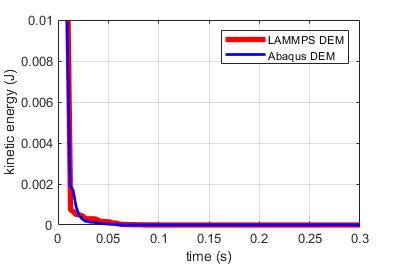}
    \end{subfigure}
    \begin{subfigure}[b]{0.49\linewidth}
        \includegraphics[width=2.4in, trim={0 0 0 0}, clip]{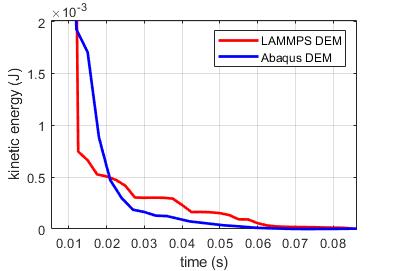}
    \end{subfigure}
    \caption{Plot of kinetic energy versus time for the entire simulation (top) and zoomed in (bottom).}
    \label{fig:kinEnergy}
\end{figure}

To strengthen the argument that Abaqus' contact performs similar to LAMMPS', a comparison can also be made using kinetic energy. For each simulation, the magnitude of the translational velocities of the impactor were exported and read into a script to calculate kinetic energy throughout the simulation. This approach allows dispersion of the impactor's kinetic energy to be plotted together, shown in Figure \ref{fig:kinEnergy}. The top plot shows the all kinetic energy calculated through the simulation and the bottom plot shows the zoomed in region between 0.01s and 0.08s where most kinetic energy dispersion into the granular bed occurs. First, the slightly varied topography of each simulation is evident by the deviation of kinetic energy trends. Since the granular beds in each simulation were generated independently under gravity, the exact `pathways' through which kinetic energy is drained are not identical during the impacts. However, in both simulations, the kinetic energy settles to zero at almost the exact same point in between 0.06s and 0.07s. This observation implies that the average rate by which kinetic energy is drained into the granular bed is very similar in both cases. Slight discrepancies noticed over the course of the dispersion are attributed to the slight difference in material property and contact definition in Abaqus and LAMMPS models. To conclude, these findings support the performance of Abaqus' contact algorithm in the discrete modeling application.

\section*{Appendix II: Plotting stress distribution throughout a bed of angular grains under surface impact}

\begin{figure}
    \centering
    \begin{subfigure}[b]{0.49\linewidth}
        \includegraphics[width=3.1in, trim={0cm 0cm 0cm 0cm}, clip]{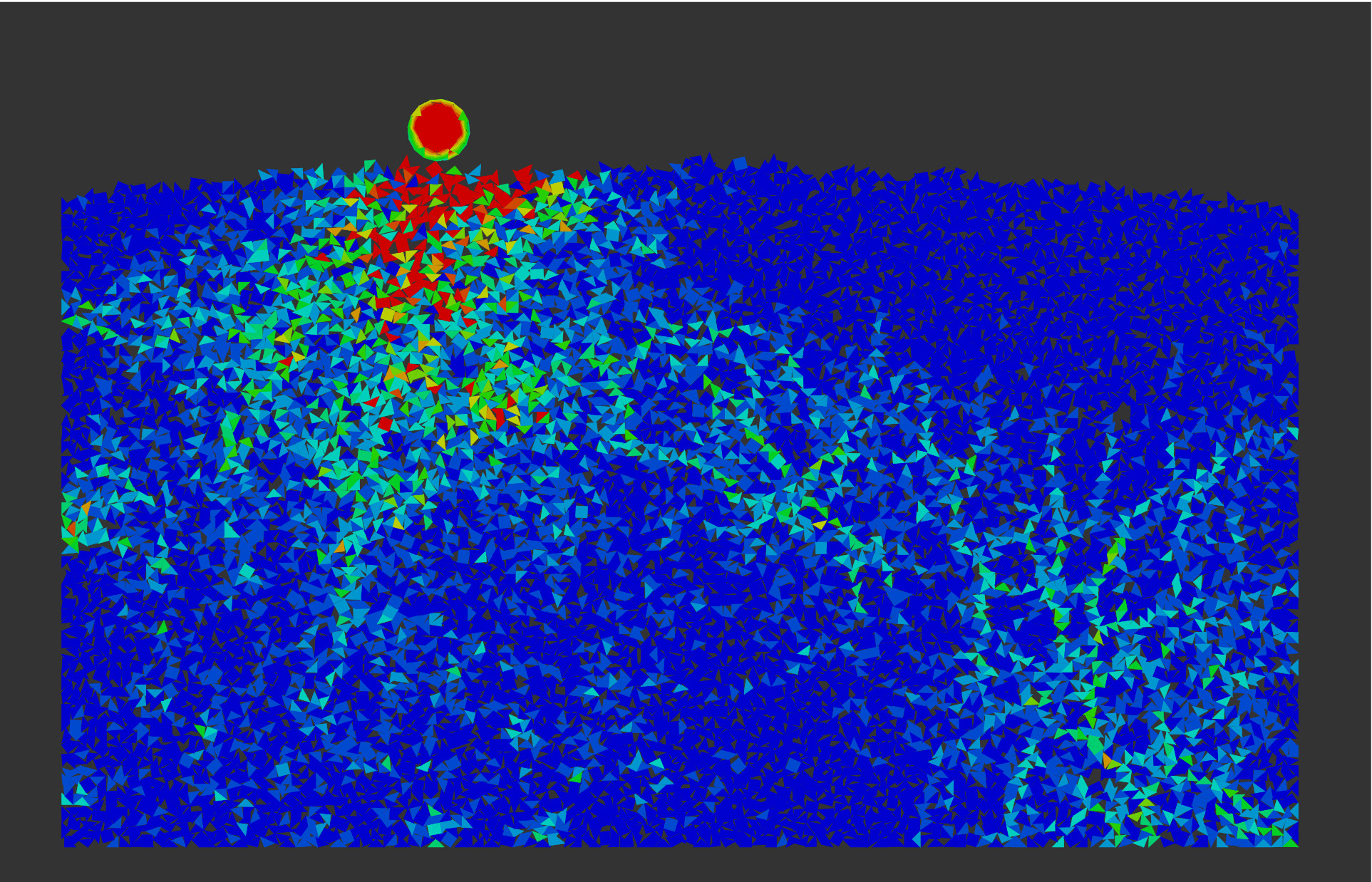}
        \subcaption{0.006s, first contact}
    \end{subfigure}
    \begin{subfigure}[b]{0.49\linewidth}
        \includegraphics[width=3.1in, trim={0cm 0cm 0cm 0cm}, clip]{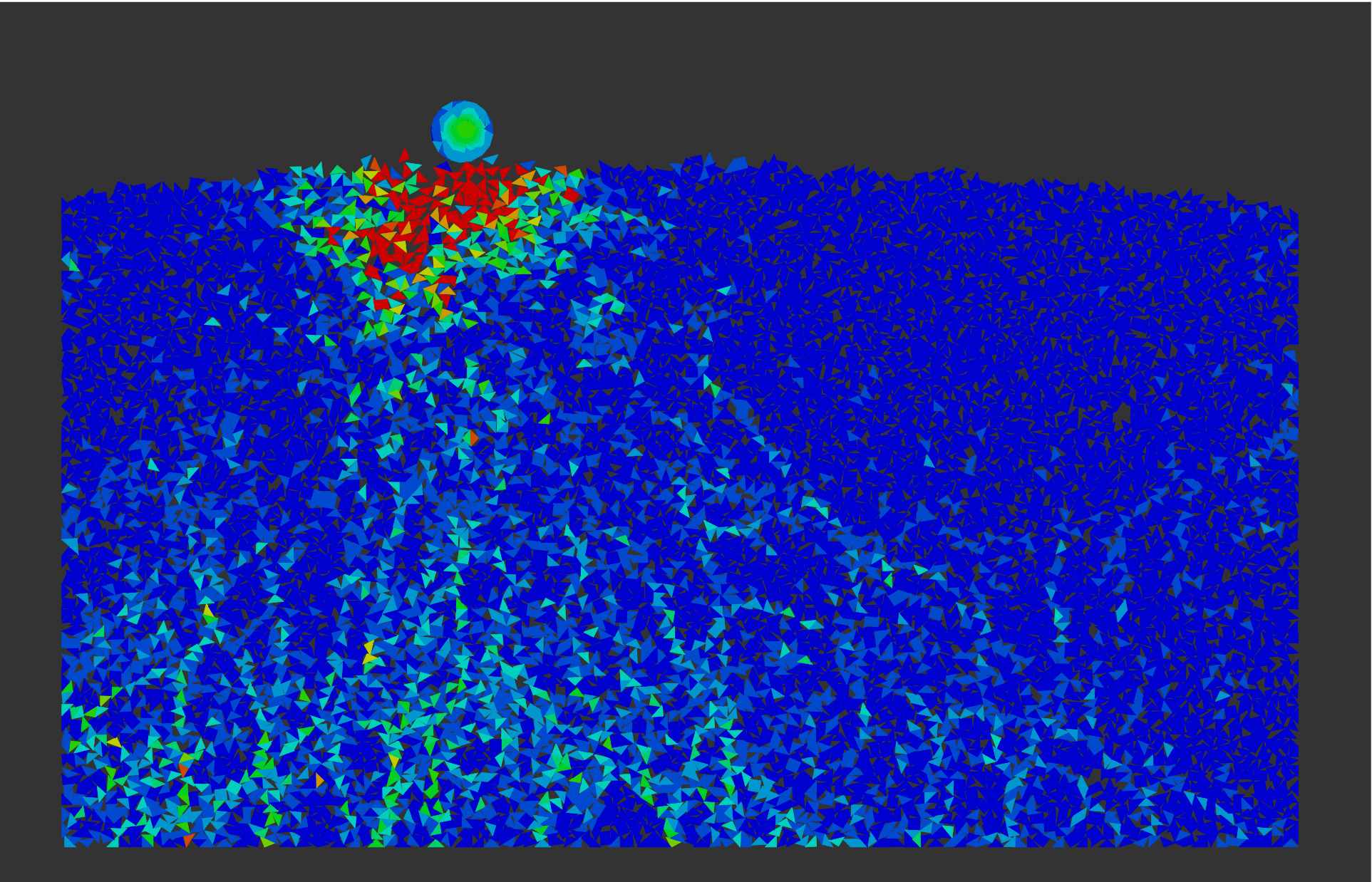}
        \subcaption{0.009s}
    \end{subfigure}
    \begin{subfigure}[b]{0.49\linewidth}
        \includegraphics[width=3.1in, trim={0cm 0cm 0cm 0cm}, clip]{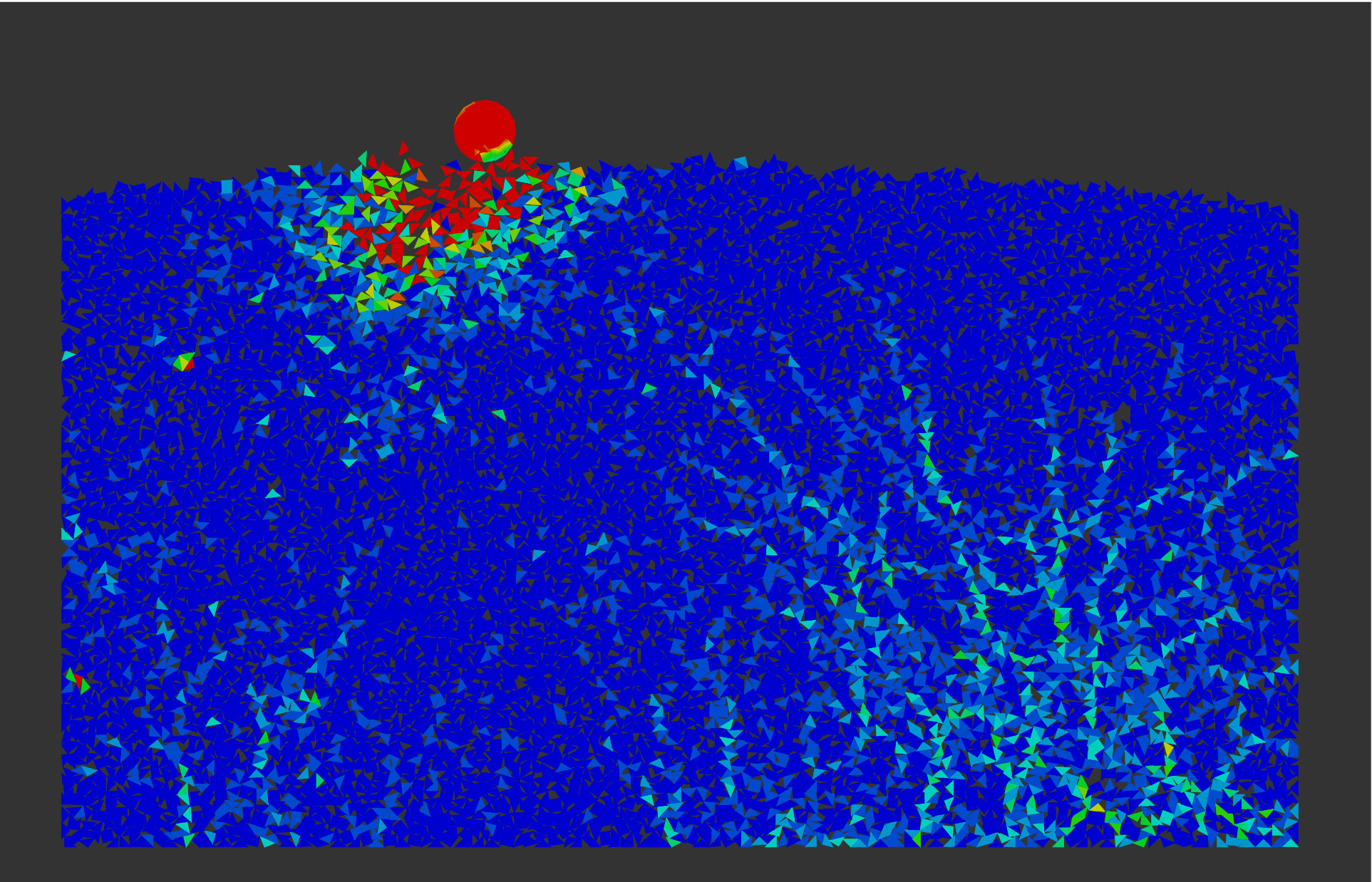}
        \subcaption{0.012s}
    \end{subfigure}
    \begin{subfigure}[b]{0.49\linewidth}
        \includegraphics[width=3.1in, trim={0cm 0cm 0cm 0cm}, clip]{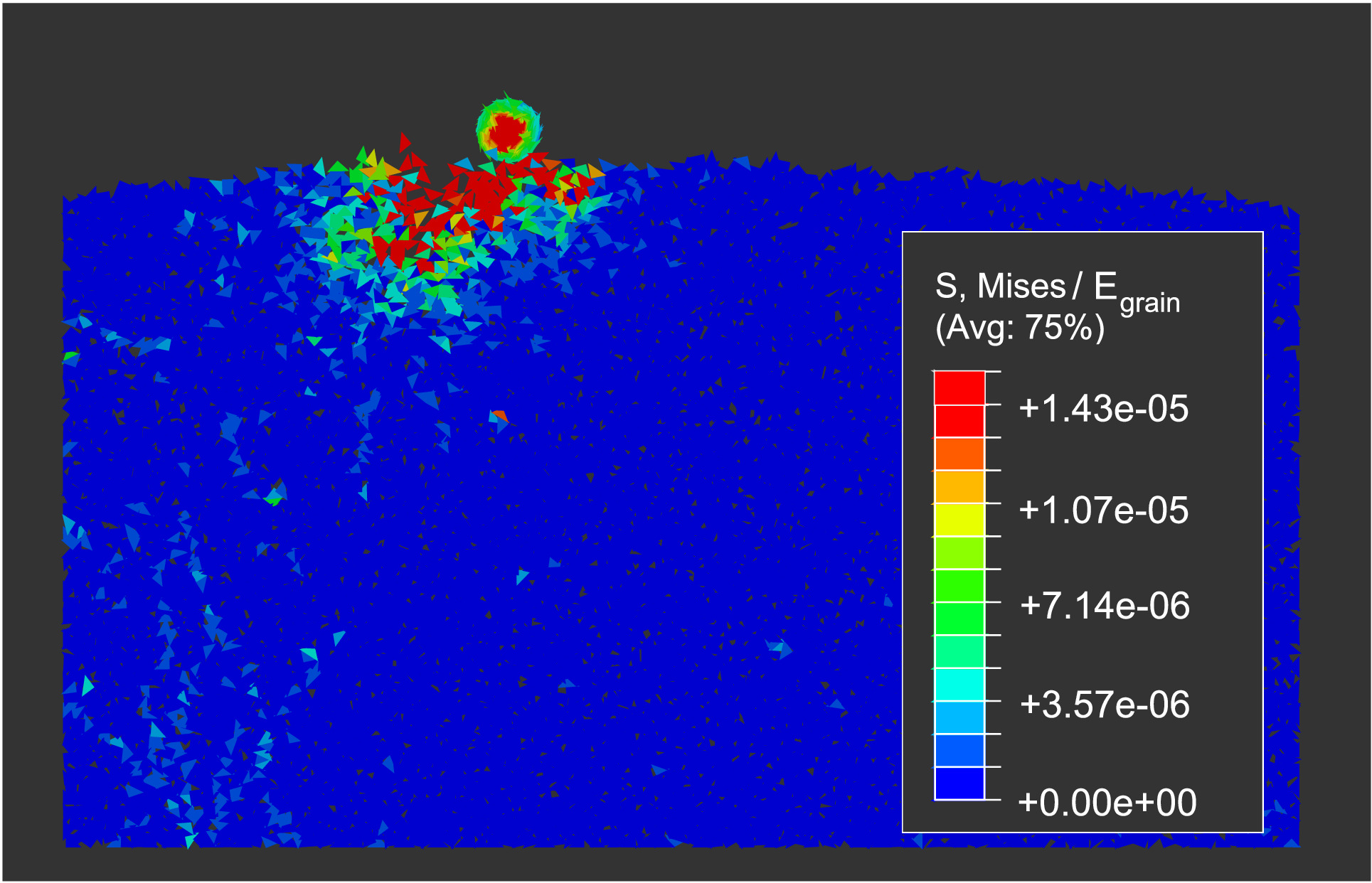}
        \subcaption{0.015s, beginning ricochet}
    \end{subfigure}
    \caption{A visualization of stress propagating throughout the granular bed under a 3.5 m/s, 45\degree\:surface impact. Reported values are von Mises stress normalized against the Young's modulus of the grain material (7.0 GPa).}
    \label{fig:stressPlots}
\end{figure}

Abaqus reports the stress experienced within each grain. We use a 3-noded linear triangular element (CPE3 triangle element) and therefore all gradient fields including stress and strain are constant within the element. This data provides a way to look at force propagation throughout the granular bed underneath an impact. Figure \ref{fig:stressPlots} shows this stress contour for several frames during an impact at 3.5 m/s and 45\degree. The von Mises stress for each element is plotted, normalized against the Young's modulus prescribed to the grains (see Table \ref{tab:props}).

Figure \ref{fig:stressPlots}a is the first frame of contact between projectile and bed. A clear pathway of forces is seen from the site of impact to the lower right corner of the bed at an angle similar to 45\degree, which corresponds to the angle of the impact. Since this impact case produces a ricochet, Figure \ref{fig:stressPlots}d shows the stress field immediately before the projectile departs the impact site. The spiderweb of forces has disappeared, though local stresses around the projectile are still high, likely driving the projectile's launch from the surface. Figure \ref{fig:stressPlots}b and Figure \ref{fig:stressPlots}c show intermediate stress distributions seen during impact.

Stress distributions around the impact site will be a central component of our future impacts work. Defining the characteristics of force chain networks and their correspondence with surface impact dynamics will undoubtedly yield a deeper understanding of low-velocity impact dynamics.

\bio{}
\endbio

\bio{}
\endbio

\end{document}